\documentclass[10pt]{article}

\usepackage{graphicx}
\usepackage{url}
\providecommand{\keywords}[1]{\textbf{\textit{Keywords---}} #1}
\usepackage[margin=1in]{geometry}
\usepackage{authblk}

\usepackage[pdftex, plainpages=false, hypertexnames=false, pdfpagelabels=true,
    hyperindex=true, linktocpage, pagebackref=true, pdfa=true]{hyperref}
    
\usepackage{fancyhdr}
\title{A Manifesto for Future Generation Cloud Computing: Research Directions for the Next Decade}

\author[1]{Rajkumar Buyya\footnote{Corresponding author; rbuyya@unimelb.edu.au}}
\author[2,1]{Satish Narayana Srirama\footnote{Co-led this work with first author; Co-First author; Corresponding author; srirama@ut.ee}}
\author[3]{Giuliano Casale}
\author[4]{Rodrigo Calheiros}
\author[5]{Yogesh Simmhan}
\author[6]{Blesson Varghese}
\author[3]{Erol Gelenbe}
\author[4]{Bahman Javadi}
\author[7]{Luis Miguel Vaquero}
\author[8]{Marco A. S. Netto}
\author[21]{Adel Nadjaran Toosi}
\author[1]{Maria Alejandra Rodriguez}
\author[9]{Ignacio M. Llorente}
\author[10]{Sabrina De Capitani di Vimercati}
\author[10]{Pierangela Samarati}
\author[11]{Dejan Milojicic}
\author[12]{Carlos Varela}
\author[13]{Rami Bahsoon}
\author[14]{Marcos Dias de Assuncao}
\author[15]{Omer Rana}
\author[16]{Wanlei Zhou}
\author[17]{Hai Jin}
\author[18]{Wolfgang Gentzsch}
\author[19]{Albert Y. Zomaya}
\author[20]{Haiying Shen}

\affil[1]{\small University of Melbourne, Australia}
\affil[2]{University of Tartu, Estonia}
\affil[3]{Imperial College London, UK}
\affil[4]{Western Sydney University, Australia}
\affil[5]{Indian Institute of Science, India}
\affil[6]{Queen's University Belfast, UK}
\affil[7]{Dyson, UK}
\affil[8]{IBM Research, Brazil}
\affil[9]{Universidad Complutense de Madrid, Spain}
\affil[10]{Universita degli Studi di Milano, Italy}
\affil[11]{Hewlett Packard Labs, USA}
\affil[12]{Rensselaer Polytechnic Institute, USA}
\affil[13]{University of Birmingham, UK}
\affil[14]{INRIA, France}
\affil[15]{Cardiff University, UK}
\affil[16]{University of Technology Sydney, Australia}
\affil[17]{Huazhong University of Science and Technology, China}
\affil[18]{UberCloud, USA}
\affil[19]{University of Sydney, Australia}
\affil[20]{University of Virginia, USA}
\affil[21]{Monash University, Australia}

\begin{document}
\maketitle

\begin{abstract}
The Cloud computing paradigm has revolutionised the computer science horizon during the past decade and has enabled the emergence of computing as the fifth utility. It has captured significant attention of academia, industries, and government bodies. Now, it has emerged as the backbone of modern economy by offering subscription-based services anytime, anywhere following a pay-as-you-go model. This has instigated (1) shorter establishment times for start-ups, (2) creation of scalable global enterprise applications, (3) better cost-to-value associativity for scientific and high performance computing applications, and (4) different invocation/execution models for pervasive and ubiquitous applications. The recent technological developments and paradigms such as serverless computing, software-defined networking, Internet of Things, and processing at network edge are creating new opportunities for Cloud computing. However, they are also posing several new challenges and creating the need for new approaches and research strategies, as well as the re-evaluation of the models that were developed to address issues such as scalability, elasticity, reliability, security, sustainability, and application models. The proposed manifesto addresses them by identifying the major open challenges in Cloud computing, emerging trends, and impact areas. It then offers research directions for the next decade, thus helping in the realisation of Future Generation Cloud Computing.

\end{abstract}


\section{Introduction}

Cloud computing has shaped the way in which software and IT infrastructure are used by consumers and triggered the emergence of computing as the fifth utility~\cite{buyya2009cloud}. Since its emergence, industry organisations, governmental institutions, and academia have embraced it and its adoption has seen a rapid growth. This paradigm has developed into the backbone of modern economy by providing on-demand access to subscription-based IT resources, resembling not only the way in which basic utility services are accessed but also the reliance of modern society on them. Cloud computing has enabled new businesses to be established in a shorter amount of time, has facilitated the expansion of enterprises across the globe, has accelerated the pace of scientific progress, and has led to the creation of various models of computation for pervasive and ubiquitous applications, among other benefits.

Up to now, there have been three main service models that have fostered the adoption of Clouds, namely Software, Platform, and Infrastructure as a Service (SaaS, PaaS, and IaaS). SaaS offers the highest level of abstraction and allows users to access applications hosted in Cloud data centres (CDC), usually over the Internet. This, for instance, has allowed businesses to access software in a flexible manner by enabling unlimited and on-demand access to a range of ready-to-use applications. SaaS has also allowed organisations to avoid incurring in internal or direct expenses, such as license fees and IT infrastructure maintenance. PaaS is tailored for users that require more control over their IT resources and offers a framework for the creation and deployment of Cloud applications that includes features such as programming models and auto-scaling. This, for example, has allowed developers to easily create applications that benefit from the elastic Cloud resource model. Finally, IaaS offers access to computing resources, usually by leasing Virtual Machines (VMs) and storage space. This layer is not only the foundation for SaaS and PaaS, but has also been the pillar of Cloud computing. It has done so by enabling users to access the IT infrastructure they require only when they need it, to adjust the amount of resources used in a flexible way, and to pay only for what has been used, all while having a high degree of control over the resources.

\subsection{Motivation and Goals of the Manifesto}

Throughout the evolution of Cloud computing and its increasing adoption, not only have the aforementioned models advanced and new ones emerged, but also the technologies in which this paradigm is based (e.g., virtualization) have continued to progress. For instance, the use of novel virtualization techniques such as containers that enable improved utilisation of the physical resources and further hide the complexities of hardware is becoming increasingly widespread, even leading to a new service model being offered by providers known as Container as a Service (CaaS). There has also been a rise in the type and number of specialised Cloud services that aid industries in creating value by being easily configured to meet specific business requirements. Examples of these are emerging, easy-to-use, Cloud-based data analytics services and serverless architectures.

Another clear trend is that Clouds are becoming increasingly geographically distributed to support emerging application paradigms. For example, Cloud providers have recently started extending their infrastructure and services to include edge devices for supporting emerging paradigms such as the Internet of Things (IoT) and Fog computing. Fog computing aims at moving decision making operations as close to the data sources as possible by leveraging resources on the edge such as mobile base stations, gateways, network switches and routers, thus reducing response time and network latencies. Additionally, as a way of fulfilling increasingly complex requirements that demand the composition of multiple services and as a way of achieving reliability and improving sustainability, services spanning across multiple geographically distributed CDCs have also become more widespread. 

\begin{figure}
  \includegraphics[width=0.8\textwidth]{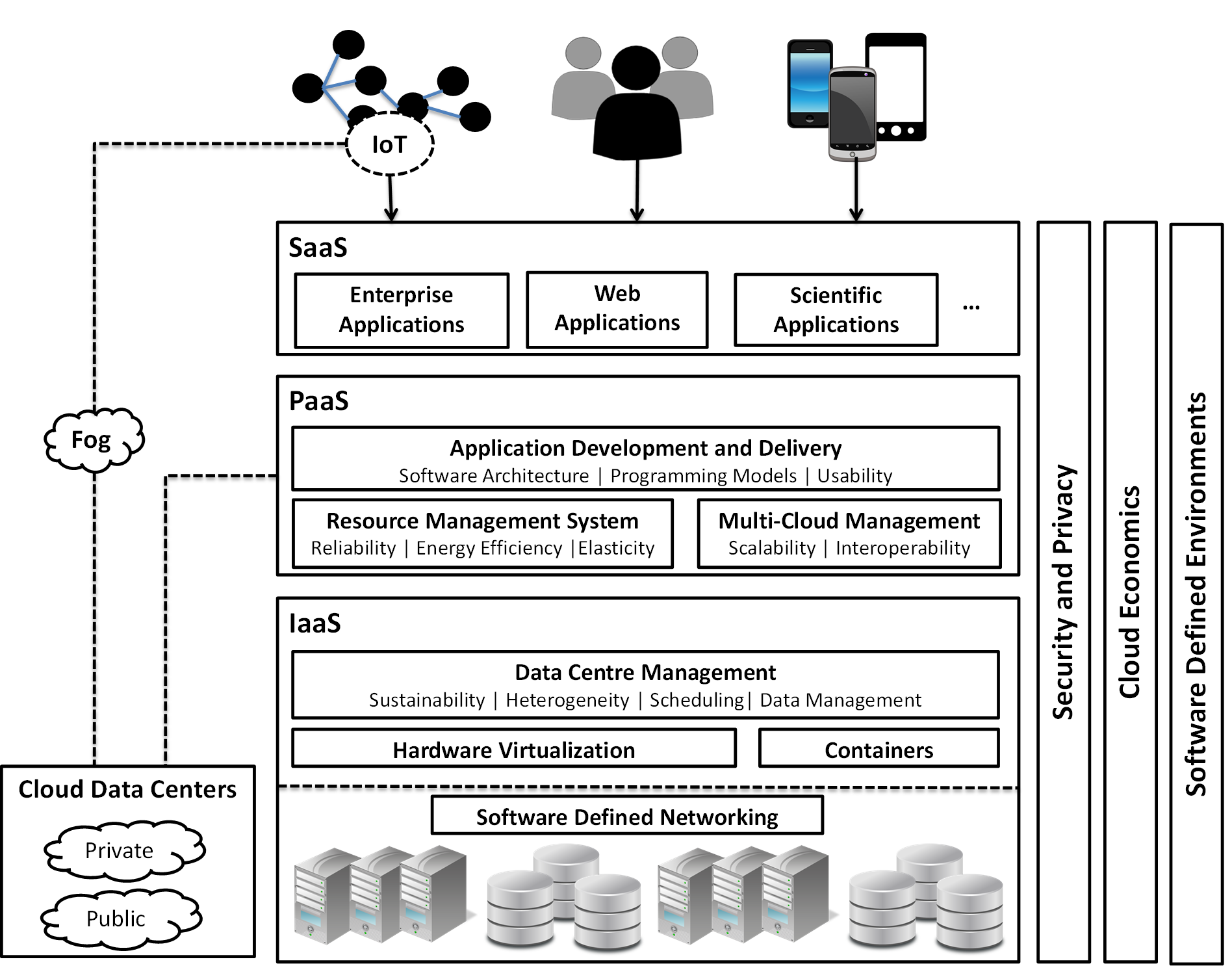} 
  \caption{Components of the Cloud computing paradigm}
  \label{fig:CloudComponents}
\end{figure}

The adoption of Cloud computing will continue to increase and support for these emerging models and services is of paramount importance. In 2016, the IDG's Cloud adoption report found that 70\% of organisations have at least one of their applications deployed in the Cloud and that the numbers are growing~\cite{IDG:CloudSurvey.16}. In the same year, the IDC's (International Data Corporation) Worldwide Semiannual Public Cloud Services Spending Guide~\cite{IDC:CloudEconomics.17} reported that Cloud services were expected to grow from \$70 billion in 2015 to more than \$203 billion in 2020, an annual growth rate almost seven times the rate of overall IT spending growth. This extensive usage of Cloud computing in various emerging domains is posing several new challenges and is forcing us to rethink the research strategies and re-evaluate the models that were developed to address issues such as scalability, resource management, reliability, and security for the realisation of next-generation Cloud computing environments~\cite{varghese2017next}.

This comprehensive manifesto brings these advancements together and identifies open challenges that need to be addressed for realising the \textit{Future Generation Cloud Computing}. Given that rapid changes in computing/IT technologies in a span of 4-5 years are common, and the focus of the manifesto is for the next decade, we envision that identified research directions get addressed and will have impact on the next two or three generations of utility-oriented Cloud computing technologies, infrastructures, and their applications' services. The manifesto first discusses major challenges in Cloud computing, investigates their state-of-the-art solutions, and identifies their limitations. The manifesto then discusses the emerging trends and impact areas, that further drive these Cloud computing challenges. Having identified these open issues, the manifesto then offers comprehensive future research directions in the Cloud computing horizon for the next decade. Figure~\ref{fig:CloudComponents} illustrates the main components of the Cloud computing paradigm and positions the identified trends and challenges, which are discussed further in the next sections.

The rest of the paper is organised as follows: Section~\ref{sec:challenges} discusses the state-of-the-art of the challenges in Cloud computing and identifies open issues. Section~\ref{sec:trends} discusses the emerging trends and impact areas related to the Cloud computing horizon. Section~\ref{sec:FutureResearch} provides a detailed discussion about the future research directions to address the open challenges of Cloud computing. In the process, the section also mentions how the respective future research directions will be guided and influenced by the emerging trends. Section~\ref{sec:summary} provides a conclusion for the manifesto. 

\section{Challenges: State-of-the-Art and Open Issues}
\label{sec:challenges}

\begin{figure}
  \includegraphics[width=\textwidth]{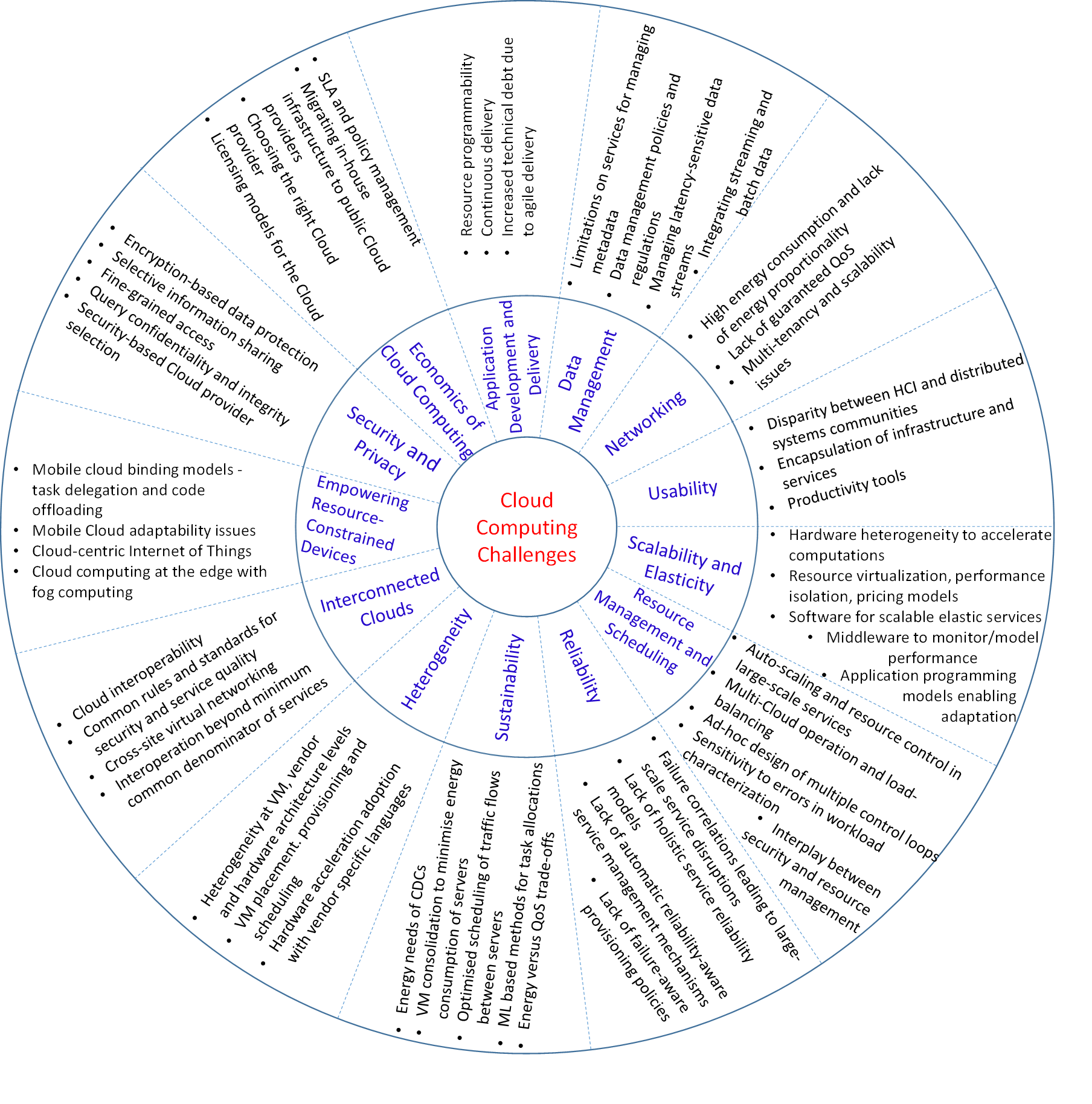}
  \caption{Cloud computing challenges, state-of-the-art and open issues}
  \label{fig:CloudSOA}
\end{figure}

As Cloud computing became popular, it has been extensively utilised in hosting a wide variety of applications. It posed several challenges (shown within the inner ring in Figure~\ref{fig:CloudSOA}) such as issues with sustainability, scalability, security, and data management among the others. Over the past decade, these challenges were systematically addressed and the state-of-the-art in Cloud computing has advanced significantly. However, there remains several issues open, as summarised in the outer ring of Figure~\ref{fig:CloudSOA}. The rest of the section identifies and details the challenges in Cloud computing and their state-of-the-art, along with the limitations driving their future research.

\subsection{Scalability and Elasticity}

Cloud computing differs from earlier models of distributed computing such as grids and clusters, in that it promises virtually unlimited computational resources on demand.  At least two clear benefits can be obtained from this promise:  first, unexpected peaks in computational demand do not entail breaking service level agreements (SLAs) due to the inability of a fixed computing infrastructure to deliver users' expected quality of service (QoS), and second, Cloud computing users do not need to make significant up-front investments in computing infrastructure but can rather grow organically as their computing needs increase and only pay for resources as needed.  The first (QoS) benefit of the Cloud computing paradigm can only be realised if the infrastructure supports {\it scalable} services, whereby additional computational resources can be allocated, and new resources have a direct, positive impact on the performance and QoS of the hosted applications.  The second (economic) benefit can only be realised if the infrastructure supports {\it elastic} services, whereby allocated computational resources can {\it follow demand} and by dynamically growing and shrinking prevent over- and under-allocation of resources.

The research challenges associated with {\it scalable services} can be broken into hardware, middleware, and application levels. Cloud computing providers must embrace parallel computing {\it hardware} including multi-core, clusters, accelerators such as Graphics Processing Units (GPUs)~\cite{younge-ipdps-2014}, and non-traditional (e.g., neuromorphic and future quantum) architectures, and they need to present such heterogeneous hardware to IaaS Cloud computing users in abstractions (e.g., VMs, containers) that while providing isolation, also enable performance guarantees.  At the {\it middleware} level, programming models and abstractions are necessary, so that PaaS Cloud computing application developers can focus on functional concerns (e.g., defining {\it map} and {\it reduce} functions) while leaving non-functional concerns (e.g., scalability, fault-tolerance) to the middleware layer~\cite{imai2016developing}.  At the {\it application} level, new generic algorithms need to be developed so that inherent scalability limitations of sequential deterministic algorithms can be overcome; these include asynchronous evolutionary algorithms, approximation algorithms, and online/incremental algorithms (see e.g.,~\cite{desell2009robust}). These algorithms may trade off precision or consistency for scalability and performance.

Ultimately, the scalability of the Cloud is limited by the extent to which individual components, namely compute, storage and interconnects scale. Computation has been limited by the end of scaling of both Moore's law (doubling the number of transistors every 1.5 year) and Dennard scaling (``the power use stays in proportion with area: both voltage and current scale (downward) with length''). As a consequence, the new computational units do not scale any more, nor does the power use scale. This directly influences the scaling of computation performance and cost of the Cloud. Research in new technologies, beyond CMOS (Complementary Metal-Oxide-Semiconductor), is necessary for further scaling.  Similar is true for memory. DRAM (Dynamic Random-Access Memory) is limiting the cost and scaling of existing computers and new non-volatile technologies are being explored that will introduce additional scaling of load-store operating memory while reducing the power consumption. Finally, the photonic interconnects are the third pillar that enables the so called silicon photonics to propagate photonic connections into the chips improving performance, increasing scale, and reducing power consumption.

On the other hand, the research challenges associated with {\it elastic services} include the ability to accurately predict computational demand and performance of applications under different resource allocations~\cite{imai2013accurate,srirama2014optimal}, the use of these workload and performance models in informing resource management decisions in middleware~\cite{imai2017maximum}, and the ability of applications to scale up and down, including dynamic creation, mobility, and garbage collection of VMs, containers, and other resource abstractions~\cite{varela-pdcs-mitpress-2013}. While virtualization (e.g., VMs) has achieved steady maturity in terms of performance guarantees rivalling native performance for CPU-intensive applications, ease of use of containers (especially quick restarts) has led to the adoption of containers by the developers community~\cite{felter-ispass-2015}. Programming models that enable dynamic reconfiguration of applications significantly help in elasticity~\cite{varela-agha-salsa-oopsla-2001}, by allowing middleware to move computations and data across Clouds, between public and private Clouds, and closer to edge resources as needed by future Cloud applications running over sensor networks such as the IoT.

In summary, scalability and elasticity provide operational capabilities to improve performance of Cloud computing applications in a cost-effective way, yet to be fully exploited. However, resource management and scheduling mechanisms need to be able to strategically use these capabilities.

\subsection{Resource Management and Scheduling}

The scale of modern CDCs has been rapidly growing and as of today they contain computing and storage devices in the range of tens to hundreds of thousands, hosting complex Cloud applications and relevant data. This makes the adoption of effective resource management and scheduling policies important to achieve high scalability and operational efficiency.

Nowadays, IaaS providers mostly rely on either {\em static} VM provisioning policies, which allocate a fixed set of physical resources to VMs using bin-packing algorithms, or {\em dynamic} policies, capable of handling load variations through live VM migrations and other load balancing techniques~\cite{mann2015allocation}. These policies can either be reactive or proactive, and typically rely on knowledge of VM resource requirements, either user-supplied or estimated using monitoring data and forecasting.

Resource management methods are also important for PaaS and SaaS providers to help managing the type and amount of resources allocated to distributed applications, containers, web-services and micro-services. Policies available at this level include for example: 1) auto-scaling techniques, which dynamically scale up and down resources based on current and forecasted workloads; 2) resource throttling methods, to handle workload bursts, trends, smooth auto-scaling transients, or control usage of preemptible VMs (e.g., micro VMs); 3) admission control methods, to handle peak load and prioritize workloads of high-value customers; 4) service orchestration and workflow schedulers, to compose and orchestrate workloads, possibly specialised for the target domain (e.g., scientific data workflows~\cite{malawski2015algorithms}), which make decisions based on their cost-awareness and the constraint requirements of tasks; 5) multi-Cloud load balancers, to spread the load of an application across multiple CDCs.

The area of resource management and scheduling has spawned a large body of research, some recent surveys include~\cite{ardagna2014quality, manvi2014resource, lopes2016taxonomy, singh2016qos}. However, several challenges and limitations still remain. For example, existing management policies tend to be intolerant to inaccurate estimates of resource requirements, calling for studying novel trade-offs between policy optimality and its robustness to inaccurate workload information~\cite{imai-patterson-varela-ccgrid-2018}. Further, demand estimation and workload prediction methods can be brittle and it remains an open question whether Machine Learning (ML) and Artificial Intelligence (AI) methods can fully address this shortcoming~\cite{casas2017pso}. Another frequent issue is that resource management policies tend to focus on optimising specific metrics and resources, often lacking a systematic approach to co-existence in the same environment of multiple control loops, to ensure fair resource access across users, and to holistically optimise across layers of the Cloud stack. Novel resource management and scheduling methods for hybrid Clouds and federated Clouds also need to be devised~\cite{imai2013accurate}. Risks related to the interplay between security and resource management are also insufficiently addressed in current research work.

\subsection{Reliability}

Reliability is another critical challenge in Cloud computing environments. Data centres hosting Cloud computing consist of highly interconnected, and interdependent systems. Because of their scale, complexity and interdependencies, Cloud computing systems face a variety of reliability related threats such as hardware failures, resource missing failures, overflow failures, network failures, timeout failures and flaws in software being triggered by environmental change. Some of these failures can escalate and devastatingly impact system operation, thus causing critical failures~\cite{gunawi2011failure}. Moreover, a cascade of failures may be triggered leading to large-scale service disruptions 
with far-reaching consequences~\cite{javadi2012failure}. As organisations are increasingly interested in adapting Cloud computing technology for applications with stringent reliability assurance and resilience requirements~\cite{sharma2016reliability}, there is an urgent demand for new ways to provision Cloud services with assured performance and resilience to deal with all types of independent and correlated failures~\cite{dean2009large}. Moreover, the mutual impact of reliability and energy efficiency of Cloud systems is one of the current research challenges~\cite{vishwanath2010characterizing}. 

Although reliability in distributed computing has been studied before~\cite{pezoa2012performance}, standard fault tolerance and reliability approaches cannot be directly applied in Cloud computing systems. The scale and expected reliability of Cloud computing are increasingly important but hard to analyse due to the range of inter-related characteristics, e.g. their massive-scale, service sharing models, wide-area network, and heterogeneous software/hardware components. Previously, independent failures have mostly been addressed separately, however, the investigation into their interplay has been completely ignored~\cite{ghosh2010end}. Furthermore, since Cloud computing is typically more service-oriented rather than resource-oriented, reliability models for traditional distributed systems cannot be directly applied to Cloud computing. So, existing state-of-the-art Cloud environments lack thorough service reliability models, automatic reliability-aware service management mechanisms, and failure-aware provisioning policies. 

\subsection{Sustainability}
\label{sec:challenges:Sustainability}

Sustainability is the greatest challenge of our century, and ICT in general \cite{gelenbe2015impact} utilises today close to 10\% of all electricity consumed world-wide, resulting in a CO$_2$ impact that is comparable to that of air-travel. In addition to the energy consumed to operate ICT systems, we know that substantial electricity is used to manufacture electronic components, and then decommission them after the end of their useful life-time; the amount of energy consumed in this process can be 4-5 fold greater than the electricity that this equipment will consume to operate during its lifetime.

CDC deployments until recently have mainly focused on high performance and have not paid enough attention to energy consumption. Thus, today a typical CDC's energy consumption is similar to that of 25,000 households~\cite{kaplan2008revolutionizing}, while the total number of operational CDCs worldwide is 8.5 million in 2017 according to IDC. Indeed, according to Greenpeace, Cloud computing worldwide consumes more energy than most countries and only the four largest economies (USA, China, Russia, and Japan) surpass Clouds in their annual electricity usage.
As the energy consumption, and the relative cost of energy in the total expenditures for the Cloud, rapidly increases, not enough research has gone into minimising the amount of energy consumed by Clouds, information systems that exploit Cloud systems, and networks~\cite{pernici2012can,buyya2010energy}.

On the other hand, networks and the Cloud also have a huge potential to save energy in many areas such as smart cities, or to be used to optimise the mix of renewable and non-renewable energy worldwide~\cite{shuja2017greening}. However, the energy consumption of Clouds cannot be viewed independently of the QoS that they provide, so that both energy and QoS must be managed in conjunction.  Indeed, for a given computer and network technology, reduced energy consumption is often coupled with a reduction of the QoS that users will experience. In some cases, such as critical or even life-threatening real-time needs, such as Cloud support of search and rescue operations, hospital operations or emergency management, a Cloud cannot choose to save energy in exchange for reduced QoS.

Current Cloud systems and efforts have in the past primarily focused on consolidation of VMs for minimising energy consumption of servers~\cite{berl2010energy}. But other elements of CDC infrastructures, such as cooling systems (close to 35\% of energy) and networks, which must be very fast and efficient, also consume significant energy that needs to be optimised by proper scheduling of the traffic flows between servers (and over high-speed networks) inside the data centre~\cite{gelenbe2012power}.

Because of multi-core architectures, novel hardware based sleep-start controls and clock speed  management techniques, the power consumption of servers increasingly depends, and in a non-linear manner, on their instantaneous workload. Thus new ML-based methods have been developed to dynamically allocate tasks to multiple servers in a CDC or in the Fog~\cite{wang2015adaptive} so that a combination of  violation of SLA, which are costly to the Cloud operator and inconvenient for the end user, and other operating costs including energy consumption, are minimised. Holistic techniques must also address  the QoS effect of networks such as packet delays on overall SLA, and the energy effects of networks for remote access to CDC~\cite{wang2016adaptive}. The purpose of these methods is to provide online automatic, or autonomic and self-aware methods to holistically manage both QoS and energy consumption of Cloud systems.

Recent work~\cite{yin2017multi} has also shown that deep learning with neural networks can be effectively applied in experimental but realistic settings so that tasks are allocated to servers in a manner that optimises a prescribed performance profile that can include execution delays, response times, system throughput, and energy consumption of the CDC. Another approach that maximises the sustainability of Cloud systems and networks involves rationing the energy supply~\cite{gelenbe2014adaptive} so that the CDC can modulate its own energy consumption and delivered QoS in response, dynamically modifying 
the processors' variable clock rates as a function of the supply of energy. It has also been suggested that different sources of renewable and non-renewable energy can be mixed~\cite{gelenbe2016energy}.

\subsection{Heterogeneity}
\label{subsec:SOA:Heterogeneity}

Public Cloud infrastructure has constantly evolved in the last decade. This is because service providers have increased their offerings while continually incorporating state-of-the-art hardware to meet customer demands and maximise performance and efficiency. This has resulted in an inherently heterogeneous Cloud with heterogeneity at three levels. 

The first is at the VM level, which is due to the organisation of homogeneous (or near homogeneous; for example, same processor family) resources in multiple ways and configurations. For example, homogeneous hardware processors with N cores can be organised as VMs with any subset or multiples of N cores. The second is at the vendor level, which is due to employing resources from multiple Cloud providers with different hypervisors or software suites. This is usually seen in multi-Cloud environments~\cite{lagar2009snowflock}. The third is at the hardware architecture level, which is due to employing both CPUs and hardware accelerators, such as GPUs and Field Programmable Gate Arrays (FPGAs)~\cite{singhal2013collaboration}.

The key challenges that arise due to heterogeneity in the Cloud are twofold. The first challenge is related to resource and workload management in heterogeneous environments. State-of-the-art in resource management focuses on static and dynamic VM placement and provisioning using global or local scheduling techniques that consider network parameters and energy consumption~\cite{crago2011heterogeneous}. Workload management is underpinned by benchmarking techniques that are used for workload placement and scheduling techniques. Current benchmarking practices are reasonably mature for the first level of heterogeneity and are developing for the second level~\cite{jennings2015resource,varghese2016cloud}. However, significant research is still required to predict workload performance given the heterogeneity at the hardware architecture level. Despite advances, research in both heterogeneous resource management and workload management on heterogeneous resources remain fragmented since they are specific to their level of heterogeneity and do not work across the VM, vendor, and hardware architecture levels. It is still challenging to obtain a general purpose Cloud platform that integrates and manages heterogeneity at all three levels. 

The second challenge is related to the development of application software that is compatible with heterogeneous resources. Currently, most accelerators require different (and sometimes vendor specific) programming languages. Software development practices for exploiting accelerators for example additionally require low level programming skills and has a significant learning curve. For example, CUDA or OpenCL are required for programming GPUs. This gap between hardware accelerators and high-level programming makes it difficult to easily adopt accelerators in Cloud software. It is recognised that abstracting hardware accelerators under middleware will reduce opportunities for optimising the source code for maximising performance. When the Cloud service offering is only the `infrastructure', the onus is on individual developers to provide source code that is targeted to the hardware environment. However, when services, such as `software' and `platforms' are offered on the Cloud, the onus is not on the developer since the aim of these services is to abstract the low-level technicalities away from the user. Therefore, it becomes necessary that the hardware is abstracted via a middleware for applications to exploit. Certainly, this comes at the expense of performance and fewer opportunities to optimise the code. Hence, there is a trade-off between performance and ease of use, when moving from VMs at the infrastructure level and on to using software and services available higher up in the computing stack. One open challenge in this area is developing software that is agnostic of the underlying hardware and can adapt based on the available hardware~\cite{kachris2016vineyard}. 

\subsection{Interconnected Clouds}

Although interconnection of Clouds was one of the earliest research problems that was identified in Cloud computing~\cite{buyya2010intercloud,rochwerger2009reservoir,bernstein2009blueprint}, Cloud interoperation continues to be an open issue since the field has rapidly evolved over the last half decade. Cloud providers and platforms still operate in silos, and their efforts for integration usually target their own portfolio of services. Cloud interoperation should be viewed as the capability of public Clouds, private Clouds, and other diverse systems to understand each other's system interfaces, configurations, forms of authentication and authorisation, data formats, and application initialisation and customisation~\cite{sotomayor2009virtual}.

Within the broader concept of interconnected Clouds, there are a number of methods that can be used to aggregate the functionalities and services of disparate Cloud providers and/or data centres. These techniques vary on who are the players that engage in the interconnections, its objectives, and the level of transparency in the aggregation of services offered to users~\cite{toosi2014interconnected}.

Existing public Cloud providers offer proprietary mechanisms for interoperation that exhibit important limitations as they are not based on standards and open-source, and they do not interoperate with other providers. Although there are multiple efforts for standardisation, such as Open Grid Forum's (OGF) Open Cloud Computing Interface (OCCI), Storage Networking Industry Association's (SNIA) Cloud Data Management Interface (CDMI), Distributed Management Task Force's (DMTF) Cloud Infrastructure Management Interface (CIMI), DMTF's Open Virtualization Format (OVF), IEEE's InterCloud and National Institute of Standards and Technology's (NIST) 
Federated Cloud, the interfaces of existing Cloud services are not standardised and different providers use different APIs, formats and contextualization mechanisms for comparable Cloud services. 

Broadly, the approaches can be classified as federated Cloud computing, if the interconnection is initiated and managed by providers (and usually transparent to users) as InterCloud or hybrid Clouds if initiated and managed by users or third parties on behalf of the users.

Federated Cloud computing is considered as the next step in the evolution of Cloud computing and an integral part of the new emerging Edge and Fog computing architectures. The federated Cloud model is gaining increasing interest in the IT market, since it can bring important benefits for companies and institutions, such as resource asset optimisation, cost savings, agile resource delivery, scalability, high availability and business continuity, and geographic dispersion~\cite{buyya2010intercloud}.

In the area of InterClouds and hybrid Clouds, Moreno et al. notice that a number of approaches were proposed to provide ``\emph{the necessary mechanisms for sharing computing, storage, and networking resources}''~\cite{moreno2012iaas}. This happens for two reasons. First, companies would like to use as much as possible of their existing in house infrastructures, for both economic and compliance reasons, and thus they should seamlessly integrate with public Cloud resources used by the company.  Second, for all the workloads that are allowed to go to Clouds or for resource needs exceeding on premise capabilities, companies are seeking to offload as much of their applications as possible to the public Clouds, driven not only by the economic benefits and shared resources, but also due to the potential freedom to choose among multiple vendors on their terms.

State-of-the-art projects such as Aneka~\cite{buyya2015multi} have developed middleware and library solutions for integration of different resources (VMs, databases, etc.). However, the problem with such approaches is that they need to operate in the lowest common denominator among the services offered by each provider, and this leads to suboptimal Cloud applications or support at specific service models. 

Regardless of the particular Cloud interconnection pattern in place, interoperability and portability have multiple aspects and relate to a number of different components in the architecture of Cloud computing and data centres, each of which needs to be considered in its own right. These include standard interfaces, portable data formats and applications, and internationally recognised standards for service quality and security.
The efficient and transparent provision, management and configuration of cross-site virtual networks to interconnect the on-premise Cloud and the external provider resources is still an important challenge that is slowing down the full adoption of this technology~\cite{huedo2017interoperable}.  

As Cloud adoption grows and more applications are moved to the Cloud, the need for satisfactory solutions is likely to grow. Challenges in this area concern how to go beyond the minimum common denominator of services when interoperating across providers (and thus enabling richer Cloud applications); how to coordinate authorisation, access, and billing across providers; and how to apply InterCloud solutions in the context of Fog computing and other emerging trends.

\subsection{Empowering Resource-Constrained Devices}
\label{subsec:SOA:Empowering}

Cloud services are relevant not only for enterprise applications, but also for the resource constrained devices and their applications. With the recent innovation and development, mobile devices such as smartphones and tablets, have achieved better CPU and memory capabilities. They also have been integrated with a wide range of hardware and sensors such as camera, GPS (Global Positioning System), accelerometer etc. In addition, with the advances in 4G, 5G, and ubiquitous WiFi, the devices have achieved significantly higher data transmission rates. This progress has led to the usage of these devices in a variety of applications such as mobile commerce, mobile social networking and location based services. While the advances in the mobiles are significant and they are also being used as service providers, they still have limited battery life and when compared to desktops have limited CPU, memory and storage capacities, for hosting/executing resource-intensive tasks/applications. These limitations can be addressed by harnessing external Cloud resources, which led to the emergence of Mobile Cloud paradigm.

Mobile Cloud has been studied extensively during the past years~\cite{dinh2013survey} and the research mainly focused at two of its binding models, the \textit{task delegation} and the \textit{mobile code offloading}~\cite{flores2014mobile}. With the task delegation approach, the mobile invokes web services from multiple Cloud providers, and thus faces issues such as Cloud interoperability and requirement of platform specific API. Task delegation is accomplished with the help of middlewares~\cite{flores2014mobile}. Mobile code offloading, on the other hand, profiles and partitions the applications, and the resource-intensive methods/operations are identified and offloaded to surrogate Cloud instances (Cloudlets/swarmlets). Typical research challenges here include developing the ideal offloading approach, identifying the resource-intensive methods, and studying ideal decision mechanisms considering both the device context (e.g. battery level and network connectivity) and Cloud context (e.g. current load on the Cloud surrogates)~\cite{flores2015mobile,zhou2015mcloud}. While applications based on task delegation are common, mobile code offloading is still facing adaptability challenges~\cite{flores2015mobile}.

Correspondingly, IoT has evolved as \textit{``web 4.0 and beyond''} and \textit{``Industry 4.0''}, where physical objects with sensing and actuator capabilities, along with the participating individuals, are connected and communicate over the Internet~\cite{srirama2017mobile}. There are predictions that billions of such devices/\textit{things} will be connected using advances in building innovative physical objects and communication protocols~\cite{evans2011internet}. Cloud primarily helps IoT by providing resources for the storage and distributed processing of the acquired sensor data, in different scenarios. While this \textit{Cloud-centric IoT} model~\cite{srirama2017mobile,gubbi2013internet} is interesting, it ends up with inherent challenges such as network latencies for scenarios with sub-second response requirements. An additional aspect that arises with IoT devices is their substantial energy consumption, which can be mitigated by the use of renewable energy~\cite{gelenbe2016energy}, but this in turn raises the issue of QoS as the renewable energy sources are generally sporadic. To address these issues and to realise the IoT scenarios, Fog computing is emerging as a new trend to bring computing and system supervisory activities closer to the IoT devices themselves, which is discussed in detail in Section~\ref{subsec:fog}. Fog computing mainly brings several advantages to IoT devices, such as security for edge devices, cognition of situations, agility of deployment, ultra-low latency, and efficiency on cost and performance, which are all critical challenges in the IoT environments.

\subsection{Security and Privacy}
\label{subsec:SOA:secandpri}

Security is a major concern in ICT systems and Cloud computing is no exception. Here, we provide an overview of the existing solutions addressing problems related to the secure and private management of data and computations in the Cloud (confidentiality, integrity, and availability) along with some observations on their limitations and challenges that still need to be addressed. 

With respect to the confidentiality, existing solutions typically encrypt the data before storing them at external Cloud providers~\cite{hacigumucs2002executing}. Encryption, however, limits the support of query evaluation at the provider side. Solutions addressing this problem include the definition of \textit{indexes\/}, which enable (partial) query evaluation at the provider side without the need to decrypt data, and the use of \textit{encryption techniques\/} that support the execution of operations or the evaluation of conditions directly over encrypted data. Indexes are metadata that preserve some of the properties of the attributes on which they have been defined and can then be used for query evaluation (e.g.,~\cite{agrawal2004order,damiani2003balancing,hacigumucs2002executing}). The definition of indexes must balance precision and privacy: precise indexes offer efficient query execution, but may lead to improper exposure of confidential information. Encryption techniques supporting the execution of operations on encrypted data without decryption are, for example, Order Preserving Encryption (OPE) that allows the evaluation of range conditions (e.g.,~\cite{agrawal2004order,wl2006}), and fully (or partial) homomorphic encryption that allows the evaluation of arbitrarily complex functions on encrypted data (e.g.,~\cite{bv2011,g2009,gsw2013}). Taking these encryption techniques as basic building blocks, some encrypted database systems have been developed (e.g.,~\cite{arasu2013orthogonal,popa2011cryptdb}), which support SQL queries over encrypted data.

Another interesting problem related to the confidentiality and privacy of data arises when considering modern Cloud-based applications (e.g., applications for accurate social services, better healthcare, detecting fraud, and national security) that explore data over multiple data sources with cross-domain knowledge. A major challenge of such applications is to preserve privacy, as data mining tools with cross-domain knowledge can reveal more personal information than anticipated, therefore prohibiting organisations to share their data. A research challenge is the design of theoretical models and practical mechanisms to preserve privacy for cross-domain knowledge~\cite{zhu2017differentially}. Furthermore, the data collected and stored in the Cloud (e.g., data about the  techniques, incentives, internal communication structures, and behaviours of attackers) can be used to verify and evaluate new theory and technical methods (e.g.,~\cite{han2017using,tari2015security}). A current booming trend is to use ML methods in information security and privacy to analyse Big Data for threat analysis, attack intelligence, virus propagation, and data correlations~\cite{han2016game}.

Many approaches protecting the confidentiality of data rely on the implicit assumption that any authorised user, who knows the decryption key, can access the whole data content. However, in many situations there is the need of supporting \textit{selective visibility} for \textit{different users}. Works addressing this problem are based on \textit{selective encryption} and on \textit{attribute-based encryption} (ABE)~\cite{vimercati2010encryption}. Policy updates are supported, for example,  by \textit{over-encryption}, which however requires the help of the Cloud provider, and by the \textit{Mix\&Slice} approach~\cite{bacis2016mix}, which departs from the support of the Cloud provider and uses different rounds of encryption to provide complete mixing of the resource. The problem of selective sharing has been considered also in scenarios where different parties cooperate for sharing data and to perform distributed computations.

Alternative solutions to encryption have been adopted when associations among the data are more sensitive than the data themselves~\cite{ciriani2010combining}. Such solutions split data in different fragments stored at different servers or guaranteed to be non linkable. They support only certain types of sensitive constraints and queries and the computational complexity for retrieving data increases.

While all solutions described above successfully provide efficient and selective access to outsourced data, they are exposed to attacks exploiting frequency of accesses to violate data and users privacy. This problem has been addressed by \textit{Private Information Retrieval} (PIR) techniques, which operate on publicly available data, and, more recently by \textit{privacy-preserving indexing techniques} based on, for example, Oblivious RAM, B-tree data structures, and binary search tree~\cite{di2016dynamic}. This field is still in its infancy and the development of practical solutions is an open problem. 

With respect to the integrity, different techniques such as digital signatures, Provable Data Possession, Proof Of Retrievability, let detecting unauthorised modifications of data stored at an external Cloud provider. Verifying the integrity of stored data by its owner and authorised users is, however, only one of the aspects of integrity. When data can change dynamically, possibly by multiple writers, and queries need to be supported, several additional problems have to be addressed. Researchers have investigated the use of authenticated data structures (\textit{deterministic} approaches) or insertion of integrity checks (\textit{probabilistic} approaches)~\cite{de2016efficient} to verify the correctness, completeness, and freshness of a computation. Both deterministic and probabilistic approaches can represent promising directions but are limited in their applicability and integrity guarantees provided.

With respect to the availability, some proposals have focused on the problem of how a user can select the services offered by a Cloud provider that match user's security and privacy requirements~\cite{dastjerdi2014compatibility}. Typically, the expected behaviours of Cloud providers are defined by SLAs stipulated between a user and the Cloud provider itself. Recent proposals have addressed the problem of exploring possible dependencies among different characteristics of the services offered by Cloud providers~\cite{de2016supporting}. 
These proposals represent only a first step in the definition of a comprehensive framework that allows users to select the Cloud provider that best fits their needs, and verifies that providers offer services fully compliant with the signed contract. 

Hardware-based techniques have also been adopted to guarantee the proper protection of sensitive data in the Cloud. Some of the most notable solutions include the \textit{ARM TrustZone\/} and the \textit{Intel Software Guard Extensions\/} (SGX) technology.  ARM TrustZone introduces several hardware-assisted security extensions to ARM processor cores and on-chip peripherals. The platform is then split into a ``secure world'' and a ``normal world'', each of which has different privileges and a controlled communication interface. The Intel SGX technology supports the creation of trusted execution environments, called {\em enclaves\/}, where sensitive data can be stored and processed.

Advanced \textit{cyberattacks} in the Cloud domain represent a serious threat that may affect the confidentiality, integrity, and availability of data and computations. In particular, Advanced Persistent Threats (APTs) deserves a particular mention. This is an emerging class of cyberattacks that are goal-oriented, highly-targeted, well-organised, well-funded, technically-advanced, stealthy, and persistent. The notorious Stuxnet, Flame, and Red October are some examples of APTs. APTs poses a severe threat to the Cloud computing domain, as APTs have special characteristics that can disable the existing defence mechanisms of Cloud computing such as antivirus, firewall, intrusion detection, and antivirus~\cite{xiao2017cloud}. Indeed, APT-based cyber breach instances and cybercrime activities have recently been on the rise, and it has been predicted that a 50\% increase in security budgets will be observed to rapidly detect and respond to them~\cite{brewer2014advanced}. In this context, enhancing the technical levels of cyber defence only is far from being enough~\cite{friedberg2015combating}. To mitigate the loss caused by APTs, a mixture of technical-driven security solutions and policy-driven security solutions must be designed. For example, data encryption can be viewed as the final layer of protection for ATP attacks. A policy to force all sensitive data to be encrypted and to stay in a ``trusted environment'' can prevent data leakage -- even if the attack can successfully penetrate into the system, all they can see is encrypted data. Another example is to utlise one-time password for strong authentication, providing better protection to clouds. 

\subsection{Economics of Cloud Computing}

Research themes in Cloud economics have centred on a number of key aspects over recent years: (1) pricing of Cloud services -- i.e. how a Cloud provider should determine and differentiate between different capabilities they offer, at different price bands and durations (e.g. micro, mini, large VM instances); (2) brokerage mechanisms that enable a user to dynamically search for Cloud services that match a given profile within a predefined budget; (3) monitoring to determine if user requirements are being met, and identifying penalty (often financial) that must be paid by a Cloud provider if values associated with pre-agreed metrics have been violated. The last of these has seen considerable work in the specification and implementation of SLAs, including implementation of specifications such as WS-Agreement~\cite{andrieux2007web}.

SLA is traditionally a business concept, as it specifies contractual financial agreements between parties who engage in business activities. Faniyi and Bahsoon~\cite{faniyi2016systematic} observed that up to three SLA parameters (performance, memory, and CPU cycle) are often used. SLA management also relates to the supply and demand of computational resources, instances and services~\cite{buyya2011sla,bonvin2011autonomic}. A related area of \textit{policy-based approaches} is also studied extensively~\cite{casalicchio2013mechanisms}. Policy-based approaches are effective when resource adaptation scenarios are limited in number. As the number of encoded policies grow, these approaches can be difficult to scale. Various optimisation strategies have been used to enable SLA and policy-based resource enforcement.

Another related aspect in Cloud economics has been an understanding of how an organisation migrates current in-house or externally hosted infrastructure to Cloud providers, involving the migration of an in-house IT department to a Cloud provider. Migration of existing services needs to take account of both social and economic aspects of how Cloud services are provisioned and subsequently used, and risk associated with uptime and availability of often business critical capability. Migrating systems management capabilities outside an organisation also has an influence on what skills need to be retained within an organisation. According to a survey by RightScale~\cite{Weins:2015}, IT departments may now be acting as potential brokers for services that are hosted, externally within a data centre. Systems management personnel may now be acting as intermediaries between internal user requests and technical staff at the CDC, whilst some companies may fully rely instead on technical staff at the data centre, completely removing the need for local personnel. This would indicate that small companies, in particular, may not need to retain IT skills for systems management and administration, instead relying on pre-agreed SLAs with CDCs. This has already changed the landscape of the potential skills base in IT companies. Many Universities also make use of Microsoft Office 365 for managing email, an activity that was closely guarded and managed by their Information Services/IT departments in the past.

The above context has also been motivated with interest in new implementation technologies such as sub-second billing made possible through container-based deployments, often also referred to as ``serverless computing'', such as in Google ``functions'', AWS Lambda, amongst others. Serverless computing is discussed further in Section~\ref{subsec:serverless}. 

Licensing is another economics-related issue, which can include annual or perpetual licensing. These can be restrictive for Cloud resources (e.g. not on-demand, limited number of cores, etc.) when dealing with the demands of large business and engineering simulations for physics, manufacturing, etc. Independent Software Vendors (ISVs) such as ANSYS, Dassault, Siemens, and COMSOL are currently investigating or already have more suitable licensing models for the Cloud, such as BYOL (bring your own license), or credits/tokens/elastic units, or fully on-demand.

Another challenge in Cloud economics is related to choosing the right Cloud provider. 
Comparing offerings between different Cloud providers is time consuming and often challenging, as providers do not use the same terminology when offering computational and storage resources, making a like-for-like comparison difficult. 
A number of commercial and research grade platforms have been proposed to investigate benefit/limits of Cloud selection, such as RightScale PlanForCloud
, CloudMarketMaker~\cite{javed2016cloud}, pricing tools from particular providers (e.g. Amazon Cost Calculator
, and SMI (Service Measurement Index) for ranking Cloud services~\cite{garg2013framework}. Such platforms focus on what the user requires and hide the internal details of the Cloud provider's resource specifications and pricing models. In addition, marketplace models are also studied where users purchase services from SaaS providers that in turn procure computing resources from either PaaS or IaaS providers~\cite{anselmi2017economics}.

\subsection{Application Development and Delivery}
\label{subsec:SOA:appdev}

Cloud computing empowers application developers with the ability to programmatically control infrastructure resources and platforms. Several benefits have emerged from this feature, such as the ability to couple the application with auto-scaling controllers and to embed in the code advanced self-* mechanisms for organising, healing, optimising, and securing the Cloud application at runtime.
 
A key benefit of \textit{resource} programmability is a looser boundary between development and operations, which results in the ability to accelerate the delivery of changes to the production environment. To support this feature, a variety of agile delivery tools and model-based orchestration languages (e.g., Terraform and OASIS TOSCA) are increasingly adopted in Cloud application delivery pipelines and DevOps methodologies~\cite{bass2015devops}. These tools help automating lifecycle management, including continuous delivery and continuous integration, application and platform configuration, and testing.

In terms of \textit{platform} programmability, separation of concerns has helped in tackling the complexity of software development for the Cloud and runtime management. For example, MapReduce enables application developers to specify functional components of their application, namely \textit{map} and \textit{reduce} functions on their data; while enabling the middleware layers to deal with non-functional concerns, such as parallelisation, data locality optimisation, and fault-tolerance. Several other programming models have emerged and are currently being investigated, to cope with the increasing heterogeneity of Cloud platforms. For example, in Edge computing, the effort to split applications relies on the developers~\cite{chun2011clonecloud}. Recent efforts in this area are also not yet fully automated~\cite{kang2017neurosurgeon}. Problems of this kind can be seen in many situations. Even though it is expected that there will be a wide variety and large number of edge devices and applications, there is a shortage of application delivery frameworks and programming models to deliver software spanning both the Edge and the CDC, to enable the use of heterogeneous hardware within Cloud applications, and to facilitate InterClouds operation.
 
Besides supporting and amplifying the above trends, an important research challenge is application evolution. Accelerated and continuous delivery may foster a short-term view of the application evolution, with a shift towards reacting to quality problems arising in production rather than avoiding them through careful design. This is in contrast with traditional approaches, where the application is carefully designed and tested to be as bug-free as possible prior to release. However, the traditional model requires more time between releases and thus it is less agile than continuous delivery methods. There is still a shortage of research in Cloud software engineering methods to combine the strengths of these two delivery approaches. For example, continuous acquisition of performance and reliability data across Cloud application releases may be used to better inform application evolution, to automate the process of identifying design anti-patterns, and to explore what-if scenario during testing of new features. Holistic methods to implement this vision need to be systematically investigated over the coming years.

\subsection{Data Management}

One of the key selling points of Cloud computing is the availability of affordable, reliable and elastic storage, that is collocated with the computational infrastructure. This offers a diverse suite of storage services to meet most common enterprise needs while leaving the management and hardware costs to the IaaS service provider. They also offer reliability and availability through multiple copies that are maintained transparently, along with disaster recovery with storage that can be replicated in different regions. A number of storage abstractions are also offered to suit a particular application's needs, with the ability to acquire just the necessary quantity and pay for it. \emph{Object-based storage} (Amazon Simple Storage Service (S3), Azure File), \emph{block storage services} (Azure Blob, Amazon Elastic Block Store (EBS)) of a disk volume, and \emph{logical HDD} (Hard Disk Drive) \emph{and SSD} (Solid-state Drive) disks that can be attached to VMs are common ones. Besides these, higher level data platforms such as NoSQL columnar databases, relational SQL databases and publish-subscribe message queues are available as well.

At the same time, there has been a proliferation of Big Data platforms~\cite{kune2016anatomy} running on distributed VM's collocated with the data storage in the data centre. The initial focus has been on batch processing and NoSQL query platforms that can handle large data volumes from web and enterprise workloads, such as Apache Hadoop, Spark and HBase. However, fast data platforms for distributed stream processing such as Apache Storm, Heron, and Apex have grown to support data from sensors and Internet-connected devices. PaaS offerings such as Amazon ElasticMR, Kinesis, Azure HDInsight and Google Dataflow are available as well.

While there has been an explosion in the data availability over the last decade, and along with the ability to store and process them on Clouds, many challenges still remain. Services for data storage have not been adequately supported by services for managing their metadata that allows data to be located and used effectively~\cite{muniswamy2010provenance}. Data security and privacy remain a concern (discussed further in Section~\ref{subsec:SOA:secandpri}), with regulatory compliance being increasingly imposed by various governments (such as the recent EU \emph{General Data Protection Regulation (GDPR) and US CLOUD Act}), as well as leakages due to poor data protection by users. Data is increasingly being sourced from the edge of the network as IoT device deployment grows, and the latency of wide area networks inhibits their low-latency processing. Edge and Fog computing may hold promise in this respect~\cite{varshney2017demystifying}.

Even within the data centre, network latencies and bandwidth between VMs, and from VM to storage can be variable, causing bottlenecks for latency-sensitive stream processing and bandwidth-sensitive batch processing platforms. Solutions such as Software Defined Networking (SDN) and Network Functions Virtualization (NFV), which can provide mechanisms required for allocating network capacity for certain data flows both within and across data centres with certain computing operations been performed in-network, are needed~\cite{liu2015cloud}. Better collocation guarantees of VMs and data storage may be required as well.

There is also increasing realisation that a lambda architecture that can process both data at rest and data at motion together is essential~\cite{kiran2015lambda}. Big Data platforms such as Apache Flink and Spark Streaming are starting to offer early solutions but further investigation is required~\cite{zhou2017knowledge}. Big Data platforms also have limited support for automated scaling out and in on elastic Clouds, and this feature is important for long-running streaming applications with dynamic workloads~\cite{kumbhare2015reactive}. While the resource management approaches discussed above can help, these are yet to be actively integrated within Big Data platforms. Fine-grained per-minute and per-second billing along with faster VM acquisition time, possibly using containers, can help shape the resource acquisition better. In addition, composing applications using serverless computing such as AWS Lambda and Azure Functions has been growing rapidly~\cite{register2018serverless}. These stateless functions can off-load the resource allocation and scaling to the Cloud platform provider while relying on external state by distributed object management services like Memcached or storage services like S3.

\subsection{Networking}

Cloud data centres are the backbone of Cloud services where application components reside and where service logic takes place for both internal and external users. Successful delivery of Cloud services requires many levels of communication happening within and across data centres. Ensuring that this communication occurs securely, seamlessly, efficiently and in a scalable manner is a vital role of the network that ties all the service components together. 

During the last decade, there has been many network-based innovations and research that have explicitly explored Cloud networking. For example, technologies such as SDN and NFV intended to build agile, flexible, and programmable computer networks to reduce both capital and operational expenditure for Cloud providers. In Section~\ref{ss:sdn} SDN and NFV are further discussed. Likewise, scaling limitations as well as the need for a flat address space and over subscription of servers also have prompted many recent advances in the network architecture such as VL2~\cite{Greenberg:2009}, PortLand~\cite{NiranjanMysore:2009}, and BCube~\cite{Guo:2009} for the CDCs. Despite all these advances, there are still many networking challenges that need to be addressed. 
 
One of the main concerns of today's CDCs is their high energy consumption. Nevertheless, the general practice in many data centres is to leave all networking devices always on~\cite{heller2010elastictree}. In addition, unlike computing servers, the majority of network elements such as switches, hubs, and routers are not designed to be energy proportional and things such as, sleeping during no traffic and adaptation of link rate during low traffic periods, are not a native part of the hardware~\cite{mahadevan2009power}. Therefore, the design and implementation of methodologies and technologies to reduce network energy consumption and make it proportional to the load remain as open challenges. 
 
Another challenge with CDC networks is related to providing guaranteed QoS. The SLAs of today's Clouds are mostly centred on computation and storage~\cite{guo2010secondnet}. No abstraction or mechanism enforcing the performance isolation and hence no SLAs beyond best effort is available to capture the network performance requirements such as delay and bandwidth guarantees. Within the data centre infrastructure, Guo et al.~\cite{guo2010secondnet} propose a network abstraction layer called VDC which works based on a source routing technique to provide bandwidth guarantees for VMs. Yet, their method does not provide any network delays guarantee. This challenge becomes even more pressing, when network connectivity must be provided over geographically distributed resources, for example, deployment of a ``virtual cluster'' spanning resources on a hybrid Cloud environment. Even though the network connectivity problem involving resources in multiple sites can be addressed using network virtualization technologies, providing performance guarantees for such networks as it traverses over the public Internet raises many significant challenges that require special consideration~\cite{toosi2014interconnected}. The primary challenge in this regard is that cloud providers do not have privileged access to the core Internet equipment as they do in their own data centres. Therefore, cloud providers' flexibility regarding routing and traffic engineering is limited to a large extent. Moreover, the performance of public network such as the Internet is much more unpredictable and changeable compared to the dedicate network of data centres which makes it more difficult to provide guaranteed performance requirements. Traditional WAN approaches such as Multi-Protocol Label Switching (MPLS) for traffic engineering in such networks are also inefficient in terms of bandwidth usage and handling latency-sensitive traffic due to lack of global view of the network~\cite{Hong:2013}. This is one of the main reasons that companies such as Google invested on its own dedicated network infrastructures to connect its data centres across the globe~\cite{jain2013b4}.

In addition, Cloud networking is not a trivial task and modern CDCs face similar challenges to building the Internet due to their size~\cite{azodolmolky2013cloud}. The highly virtualized environment of a CDC is also posing issues that have always existed within network apart from new challenges of these multi-tenant platforms. For example in terms of scalability, VLANs (Virtual Local Area Network) are a simple example. At present, VLANs are theoretically limited to 4,096 segments. Thus, the scale is limited to approximately 4,000 tenants in a multitenant environment. VXLAN offers encapsulation methods to address the limited number of VLANs. However, it is limited in multicasting, and supports Layer 2 only within the logical network. IPv4 is another example, where some Cloud providers such as Microsoft Azure admitted that they ran out of addresses. To overcome this issue the transition to the impending IPv6 adoption must be accelerated. This requirement means that the need for network technologies offering high performance, robustness, reliability, flexibility, scalability, and security never ends~\cite{azodolmolky2013cloud}.

\subsection{Usability}

The Human Computer Interface and Distributed Systems communities are still far from one another. Cloud computing, in particular, would benefit from a closer alignment of these two communities. Although much effort has happened on resource management and back-end related issues, usability is a key aspect to reduce costs of organisations exploring Cloud services and infrastructure. This reduction is possible, mainly due to labour related expenses as users can have better quality of service and enhance their productivity. The usability of Cloud~\cite{faisal2011issues} has already been identified as a key concern by NIST as described in their Cloud Usability Framework~\cite{stanton2015framework}, which highlights five aspects: capable, personal, reliable, secure, and valuable. Capable is related to meeting Cloud consumers expectations with regard to Cloud service capabilities. Personal aims at allowing users and organizations to change the look and feel of user interfaces and to customise service functionalities. Reliable, secure, and valuable are aspects related to having a system that performs its functions under state conditions, safely/protected, and that returns value to users respectively. Coupa's white paper~\cite{Coupa:Usability.12} on usability of Cloud applications also explores similar aspects, highlighting the importance of usability when offering services in the Internet.

For usability, current efforts in Cloud have mostly focused on encapsulating complex services into APIs to be easily consumed by users. One area where this is clearly visible is High Performance Computing (HPC) Cloud~\cite{netto2018hpccloud}. Researchers have been creating services to expose HPC applications to simplify their consumptions~\cite{huang2014development,church2015exposing}. These applications are not only encapsulated as services, but also receive Web portals to specify application parameters and manage input and output files.

Another direction related to usability of Cloud that got traction in the last years is DevOps~\cite{balalaie2016microservices,rajkumar2016devops}. 
Its goal is to integrate development (Dev) and operations (Ops) thus aiding faster software delivery (as also discussed in Sections \ref{subsec:SOA:appdev} and \ref{subsec:future:appdev}). DevOps has improved the productivity of developers and operators when creating and deploying solutions in Cloud environments. It is relevant not only to build new solutions in the Cloud but also to simplify the migration of legacy software from on-premise environments to multi-tenancy elastic Cloud services.

\section{Emerging Trends and Impact Areas}
\label{sec:trends}

As Cloud computing and relevant research matured over the years, it led to several advancements in the underlying technologies such as containers and software defined networks. These developments in turn have led to several emerging trends in Cloud computing such as Fog computing, serverless computing, and software defined computing. In addition to them, other emerging trends in ICT such as Big Data, machine/deep learning, and blockchain technology also have started influencing the Cloud computing research and have offered wide opportunities to deal with the open issues in Cloud-related challenges. 
Here, we discuss the emerging trends and impact areas relevant in the Cloud horizon. 

\subsection{Containers}

With the birth of Docker~\cite{merkel2014docker}, container technologies have aroused wide interest in both academia and industry~\cite{soltesz2007container}. Containers provide a lightweight environment for the deployment applications; they are stand-alone, self-contained units that package software and its dependencies together. Similar to VMs, containers enable the resources of a single compute node to be shared by enabling applications to run as isolated user space processes.

Containers rely on modern Linux operating systems' kernel facilities such as cgroups, LXC (Linux containers) and libcontainer. Docker uses Linux kernel's cgroups and namespaces to run independent ``containers'' within a physical machine. Control Groups (cgroups) provide isolation of resources such as CPU, memory, block I/O and network. On the other hand, namespaces isolate an application's view of the operating environment, that includes process trees, network, user IDs and mounted file systems. Docker contains the libcontainer library as a container reference implementation. By packing the application and related dependencies into a Docker image, Docker simplifies the deployment of the application and improves the development efficiency.

More and more Internet companies are adopting this technology and containers have become the de-facto standard for creating, publishing, and running applications. This increased demand has led for instance to the emergence of CaaS (\textit{container as a service}), a model derived from the traditional Cloud computing~\cite{ruan2016performance}. An example of this type of service is UberCloud~\cite{UberCloud.17,gentzsch2016novel}; a platform offering application containers and their execution for a variety of engineering simulations.

The increase in popularity of containers may be attributed to two main features. First, they start up very quickly and their launching time is less than a second. Second, containers have small memory footprint and consume a very small amount of resources. Compared with VMs, using containers not only improves the performance of applications, but also allows the host to support more instances simultaneously.

Despite these advantages, there are still drawbacks and challenges that need to be addressed. First, due to the sharing of the kernel, the isolation and security of containers is weaker than in VMs~\cite{xavier2013performance}, which stimulates much interest and enthusiasm of researchers. There are two promising solutions to this problem. One is to leverage new hardware features, such as the trusted execution support of Intel SGX~\cite{arnautov2016scone}. The other one is to use Unikernel, which is a kind of library operating system~\cite{Unikernel.17}. Second, trying to optimise the performance of containers is an everlasting theme. For example, to accelerate the container start-up, Slack is proposed to optimise the storage driver~\cite{harter2016slacker}. Last but not least, the management of container clusters based on users' QoS requirements is attracting significant attention. Systems for container cluster management such as Kubernetes~\cite{Kubernetes:2018}, Mesos~\cite{hindman2011mesos} and Swarm~\cite{Swarm:2018} are emerging as the core software of the Cloud computing platform.

\subsection{Fog Computing}
\label{subsec:fog}

The Fog is an extension to the traditional Cloud computing model in that the edge of the network is included in the computing ecosystem to facilitate decision making as close as possible to the data source~\cite{bonomi2012fog,vaquero2014finding,garcia2015edge}. 
The vision of Fog computing is three fold. First, to enable general purpose computing on traffic routing nodes, such as mobile base stations, gateways and routers. Second, to add compute capabilities to traffic routing nodes so as to process data as it is transmitted between user devices and a CDC. Third, to use a combination of the former. 

There are a number of benefits in using such a compute model. For example, latencies between users and servers can be reduced. Moreover, location awareness can be taken into account for geo-distributed computing on edge nodes. The Fog model inherently lends itself to improving the QoS of streaming and real-time applications. Additionally, mobility can be seamlessly supported, wireless access between user devices and compute servers can be enabled and scalable control systems can be orchestrated. These benefits make it an appropriate solution for the upcoming IoT class of applications~\cite{wang2017enorm,varshney2017demystifying,dastjerdi2016fog}.

Edge and Fog computing are normally used interchangeably, however, they are slightly different, both paradigms rely on local processing power near data sources. In Edge computing, the processing power is given to the IoT device itself, while in the Fog computing, computing nodes (e.g., Dockers and VMs) are placed very close the source of data. The Edge computing paradigm depends on how IoT devices can be programmed to interact with each other and run user defined codes. Unfortunately, standard APIs that provide such functionality are not fully adopted by current IoT sensors/actuators, and thus Fog computing seems to be the only viable/generic solutions to date~\cite{nan2016cost}.

The Fog would offer a full-stack of IaaS, PaaS, and SaaS resources, albeit not to the full extent as a CDC. Given that a major benefit of the Fog is its closer network proximity to the consumers of the services to reduce latency, it is anticipated that there will be a few Fog data centres per city. But as yet, the business model is evolving and possible locations for Fog resources range from a local coffee shop to mobile cell towers (as in Mobile Edge computing~\cite{wang2015dynamic}). Additionally, infrastructure provided by traditional private Cloud and independent Fog providers may be employed~\cite{ChangETAL:indieFog.17}. Economics related research challenges and opportunities for Fog computing are discussed further in Section~\ref{subsec:future:economics}. Although the concept of Mobile Edge computing is similar to the premise of Fog computing, it is based on the mobile cellular network and does not extend to other traffic routing nodes along the path data travels between the user and the CDC.

Advantages of Fog computing include the vertical scaling of applications across different computing tiers. This allows for example, pre-processing the data contained in packets so that value is added to the data and only essential traffic is transmitted to a CDC. 
Workloads can be (1) decomposed on CDCs and offloaded on to edge nodes, (2) migrated from a collection of user devices on to edge nodes, or (3) aggregated from multiple sensors or devices on an edge node. In the Fog layer, workloads may be deployed via containers in lieu of VMs that require more resources~\cite{liyanage2016mepaas,kang2017neurosurgeon}.

Cloud vendors have started to use edge locations to deliver security services (AWS Shield, Web Application Firewall Service) closer to users or to modify network traffic (e.g. Lambda@Edge
). Cloud providers are also asking customers to deploy on-premise storage and compute capabilities working with the same APIs as the ones they use in their Cloud infrastructure. These have made it possible to deliver the advantages of Fog architectures to the end users. For instance, in Intensive Care Units, in order to guarantee uninterrupted care when faced with a major IT outage, or to bring storage and computing capabilities to poorly connected areas (e.g. AWS Snowball Edge for the US Department of Defense
).

Other applications that can benefit from the Fog include smart city and IoT applications that are fast growing. Here, multi-dimensional data, such as text, audio and video are captured from urban and social sensors, and deep-learning models may be trained and perform inferencing to drive real-time decisions such as traffic signalling. Autonomous vehicles such as driverless cars and drones can also benefit from the processing capabilities offered by the Fog, well beyond what is hosted in the vehicle. The Fog can also offer computing and data archival capabilities. Immersive environments such as MMORPG gaming, 3D environment such as HoloLens and Google Glass, and even robotic surgery can benefit from GPGPUs that may be hosted on the Fog.

Many works such as Shi and Dustdar~\cite{shi2016promise}, Varghese et al.~\cite{varghese2016challenges}, Chang et al.~\cite{ChangETAL:indieFog.17} and Garcia Lopez et al.~\cite{garcia2015edge} have highlighted several challenges in Edge/Fog computing. Two prominent challenges that need to be addressed to enhance utility of Fog computing are mentioned here. First, tackling the complex management issues related to multi-party SLAs. To this end, as a first step responsibilities of all parties will need to be articulated. This will be essential for developing a unified and interoperable platform for management since Edge nodes are likely to be owned by different organisations. The EdgeX Foundry~\cite{EdgeXFoundry:2018} project aims to tackle some of these challenges. Second, given the possibility of multiple node interactions between a user device and CDC, security will need to be enhanced and privacy issues will need to be debated and addressed~\cite{stojmenovic2016overview}. The Open Fog consortium~\cite{openfogconsortium:2018} is a first step in this direction.

\subsection{Big Data}
\label{ss:bigdata}

There is a rapid escalation in the generation of streaming data from physical and crowd-sourced sensors as deployments of IoT, Cyber Physical Systems (CPS)~\cite{wolf2009cyber}, and micro-messaging social networks such as Twitter. This quantity is bound to grow many-fold, and may dwarf the size of data present on the public WWW, enterprises and mobile Clouds. Fast data platforms to deal with data velocity may usurp the current focus on data volume.

This has also seen the rise of in-memory and stream computation platforms such as Spark Streaming, Flink and Kafka that process the data in-memory as events or micro-batches and over the network rather than write to disk like Hadoop~\cite{zaharia2012resilient}. This offers a faster response for continuously arriving data, while also balancing throughput. This may put pressure on memory allocation for VMs, with SSD's playing a greater role in the storage hierarchy.

We are also seeing data acquisition at the edge by IoT and Smart City applications with an inherent feedback loop back to the edge. Video data from millions of cameras from city surveillance, self-driving cars, and drones at the edge is also poised to grow~\cite{satyanarayanan2015edge}. This makes latency and bandwidth between Edge and Cloud a constraint if purely performing analytics on the Cloud. Edge/Fog computing is starting to complement Cloud computing as a first-class platform, with Cloud providers already offering SDK's to make this easier from user-managed edge devices. While smartphones have already propagated mobile Clouds where applications cooperatively work with Cloud services, there will be a greater need to combine peer-to-peer computing on the Edge with Cloud services, possibly across data centres. This may also drive the need for more regional data centres to lower the network latency from the edge, and spur the growth of Fog computing.

Unlike structured data warehouses, the growing trend of \textit{``Data Lakes''} encourages enterprises to put all their data into Cloud storage, such as HDFS, to allow intelligence to be mined from it~\cite{stein2014enterprise}. However, a lack of tracking metadata describing the source and provenance of the data makes it challenging to use them, effectively forming \textit{``data graveyards''}. Many of these datasets are also related to each other through logical relationships or by capturing physical infrastructure, though the linked nature of the datasets may not be explicitly captured in the storage model~\cite{bizer2009linked}. There is heightened interest in both deep learning platforms like TensorFlow to mine such large unstructured data lakes, as well as distributed graph databases like Titan and Neo4J to explore such linked data.

\subsection{Serverless Computing}
\label{subsec:serverless}

Serverless computing is an emerging architectural pattern that changes dramatically the way Cloud applications are designed. Unlike a traditional three-tiered Cloud application in which both the application logic and the database server reside in the Cloud, in a serverless application the business logic is moved to the client; this may be embedded in a mobile app or ran on temporarily provisioned resources during the duration of the request. 
This translates to the fact that a client does not need to rent resources, for example Cloud VMs for running the server of an application~\cite{AWS:Lambda.2015}. 
This computing model implicitly handles the challenges of deploying applications on a VM, such as over/under provisioning Cloud VMs for the application, balancing the workload across the resources and ensuring reliability and fault-tolerance. In this case, the actual server is made abstract, such that properties like control, cost and flexibility, which are not conventionally considered are taken into account.

Consequently, serverless computing reduces the amount of backend code, developers need to write, and also reduces administration on Cloud resources. It appears in two forms; Backend as a Service (BaaS) and Functions as a Service (FaaS)~\cite{Roberts:Serverless.17}. This architecture is currently supported on platforms such as AWS Lambda
, IBM OpenWhisk
 and Google Cloud Functions.

It is worth noting the term ``serverless'' may be somehow misleading: it does not mean that the application runs without servers; instead, it means that the resources used by the application are managed by the Cloud provider~\cite{baldini2017serverless}. In BaaS, the server-side logic is replaced by different Cloud services that carry out the relevant tasks (for example, authentication, database access, messaging, etc.), whereas in FaaS ephemeral computing resources are utilised that are charged per access (rather than on the basis of time, which is typical of IaaS solutions).

FaaS poses new challenges particularly for resource management in Clouds that will need to be addressed. This is because arbitrary code (the function) will need to execute in the Cloud without any explicit specification of resources required for the operation. To make this possible, FaaS providers pose many restrictions about what functions can do and for how long they can operate~\cite{baldini2017serverless}. For example, they enforce limits on the amount of time a function can execute, how functions can be written (enforcing stateless computations), and how the code is deployed~\cite{baldini2017serverless}. This is restrictive in the types of applications that can make use of current FaaS models. 

The above results in new challenges from a Software Engineering perspective: applications need to be redesigned to leverage the model, forcing software engineers to shift the way they design and think about the logic of their applications. Although some of these changes, for example, making applications stateless, is also desirable if other benefits from Clouds as elasticity are to be fully leveraged at application model, there are at least two other challenges that are particularly relevant to this model, namely event-based and timeout-aware application logic. 
The former issue arises because each function can be seen as a particular response to an event that will trigger other events in response to its execution. The latter arises because serverless offers implement time-outs in their logic, so it is important that this is taken into consideration during the design and execution of functions, and strategies to circumvent the time limit of applications need to be adopted whenever it is necessary.

A full-fledged general-purpose serverless computing model is still a vision that needs to be achieved. Upcoming research has explored applications that can benefit from serverless computing~\cite{yan2016building} and platforms that match services offered by providers~\cite{hendrickson2016serverless,spillner2017snafu,mcgrath2017serverless}. As discussed by Hendrickson et al.~\cite{hendrickson2016serverless}, there are still a number of issues at the middleware layer that need to be addressed that are orthogonal to advances in the area of Cloud computing that are also necessary to better support this model. Despite these challenges, this is a promising area to be explored with significant practical and economic impact. It is predicted by Forbes that there will be a likely increase of serverless computing since a large number of `things' will be connected to the edge and data centres~\cite{Diaz:Serverless.16}.

\subsection{Software-defined Cloud Computing}
\label{ss:sdn}

Software-defined Cloud Computing is a method for the optimisation and automation of configuration process and physical resources abstraction, by extending the concept of virtualization to all resources in a data centre including compute, storage, and network~\cite{buyya2014software}. Virtualization technologies aim to mask, abstract and transparently leverage underlying resources without applications and clients having to understand physical attributes of the resource. Virtualization technologies for computing and storage resources are quite advanced to a large extent. The emerging trends in this space are the virtualization in networking aspects of Cloud, namely Software-defined networking (SDN) and Network functions virtualization (NFV).

The main motivation for SDN, an emerging networking paradigm, is due to the demand/need for agile and cost-efficient computer networks that can also support multi-tenancy~\cite{nadeau2013sdn}. SDN aims at overcoming the limitations of traditional networks, in particular networking challenges of multi-tenant environments such as CDCs where computing, storage, and network resources must be offered in slices that are independent or isolated from one another.  Early supporters of SDN were among those believing that networking equipment manufacturers were not meeting their needs particularly in terms of innovation and the development of required features of data centres. There were another group of supporters who aimed at running their network by harnessing the low-cost processing power of commodity hardware. 

SDN decouples the data forwarding functions and network control plane, which enables the network to become centrally manageable and programmable~\cite{ONF.17}. This separation offers the flexibility of running some form of logically centralised network orchestration via the software called SDN controller. The SDN controller provides vendor-neutral open standards which abstract the underlying infrastructure for the applications and network services and facilitates communication between applications wishing to interact with network elements and vice versa~\cite{nadeau2013sdn}. OpenFlow~\cite{mckeown2008openflow} is the de-facto standard of SDN and is used by most of SDN Controllers as southbound APIs for communications with network elements such as switches and routers.
 
NFV is another trend in networking which is quickly gaining attention with more or less similar goals to SDN. The main aim of NFV is to transfer network functions such as intrusion detection, load balancing, firewalling, network address translation (NAT), domain name service (DNS), to name a few, from proprietary hardware appliances to software-based applications executing on commercial off-the-shelf (COTS) equipment. NFV intends to reduce cost and increase elasticity of network functions by building network function blocks that connect or chain together to build communication services~\cite{chiosi2012network}. Han et al.~\cite{Han2015} presented a comprehensive survey of key challenges and technical requirements of NFV. Network service chaining, also known as service function chaining (SFC), is an automated process used by network operators to set up a chain of connected network services. SFC enables the assembly of the chain of virtual network functions (VNFs) in an NFV environment using instantiation of software-only services running on commodity hardware. Management and orchestration (MANO) of NFV environments is another popular research topic and a widely studied problem in the literature~\cite{Medhat2017}.

Apart from networking challenges, SDN and NFV can serve as building blocks of next-generation Clouds by facilitating the way challenges such as sustainability, interconnected Clouds, and security can be addressed. Heller et al.~\cite{heller2010elastictree} conducted one of the early attempts towards sustainability of Cloud networks using OpenFlow switches and providing network energy proportionality. The main advantage of using NFV is that Cloud service providers can launch new network function services in a more agile and flexible way.  In view of that, Eramo et al.~\cite{eramo2017approach} proposed a consolidation algorithm based on a migration policy of virtualized network function instances to reduce energy consumption. Google adopted SDN in its B4 network to interconnect its CDC with a globally-deployed software defined WAN~\cite{jain2013b4}. Yan et al.~\cite{yan2016software} investigated how SDN-enabled Cloud brings us new opportunities for tackling distributed denial-of-service (DDoS) attacks in Cloud computing environments.

\subsection{Blockchain}
\label{sec:blockchain}

In several industries, blockchain technology~\cite{swan2015blockchain} is becoming fundamental to accelerate and optimise transactions by increasing their level of traceability, reliability, and auditability. Blockchain consists of a distributed immutable ledger deployed in a decentralised network that relies on cryptography to meet security constraints~\cite{tosh2017security}. Different parties of a chain have the same copy of the ledger and have to agree on transactions being placed into the blockchain. Cloud computing is essential for blockchain as it can host not only the blockchain nodes, but services created to leverage this infrastructure. Cloud can encapsulate blockchain services in both PaaS and SaaS to facilitate their usage. This will involve also challenges related to scalability as these chains start to grow as technology matures. Cloud plays a key role in the widespread adoption of blockchain with its flexibility for dynamically allocating computing resources and managing storage~\cite{CSCC:ArchitectureBlockchain.17}. An important component of blockchain is to serve as a platform to run analytics on transaction data, which can be mixed with data coming from other sources such as IoT, financial, and weather-related services. There are many transactions that happen outside the Cloud and blockchain will force such transactions to be moved to the Cloud, which will require data centres to handle a much larger load than they currently do---thus raising issues related to sustainability, mainly in terms of infrastructure energy consumption (see Section 4.4). Such a load will come not only for the transactions themselves, but all analytics services that will benefit from this transactional data. Therefore, the difficult aspect in the Cloud to handle blockchain services comes from the need of much more efficient infrastructure for transactions and all associated dynamic computing demand from smart contracts and analytics that emerge at different times and geographies according to the transactional flows. 

Another side of blockchain and Cloud is to consider the direction where the advances in blockchain will assist Cloud computing~\cite{bahga2016blockchain,gaetani2017blockchain}. It is well known that Cloud is an important platform for collaboration and data exchange.  Blockchain can assist Cloud by creating more secure and auditable transaction platform. This is essential for several industries including health, agriculture, manufacturing, and petroleum. This is tied to the importance of data for machine learning and deep learning solutions. Such data is generated by several users and companies that want to receive profit for their data to artificial intelligence services. Blockchain can be interleaved with cloud platforms to create trusted and verifiable data marketplaces. Consequently, users and companies can trade data and insights in an efficient, reliable, and auditable fashion. The challenge in this research area involves scalability, mechanisms to verify the usefulness/quality of data, and usability tools to facilitate such blockchain-aware data trading mechanisms.

\subsection{Machine and Deep Learning}
\label{sec:deeplearning}

Due to the vast amount of data generated in the last years and the computing power increase, mainly of GPUs, AI has gained a lot of attention lately. Algorithms and models for machine learning and deep learning are relevant for Cloud computing researchers and practitioners. From one side, Cloud can benefit from machine/deep learning in order to have more optimised resource management, and on the other side, Cloud is an essential platform to host machine/deep learning services due to its pay-as-you-go model and easy access to computing resources. 

In the early 2000s, autonomic computing was a subject of study to make computing systems more efficient through automation~\cite{kephart2003vision}. There, systems would have four major characteristics: self-configuration, self-optimisation, self-healing, and self-protection. The vision may become possible with the assistance of breakthroughs in artificial intelligence and data availability. For Cloud, this means efficient ways of managing user workloads, predictions of demands for computing power, estimations of SLA violations, better job placement decisions, among others. Simplifying the selection of Cloud instances~\cite{samreen2016daleel} or optimising resource selection~\cite{bankole2013predicting} are well known examples of the use of machine learning for better use of Cloud services and infrastructure. The industry has already started to deliver auto-tuning techniques for many Cloud services so that many aspects of running the application stack are delegated to the Cloud platform. For instance, Azure SQL database has auto-tuning as a built-in feature that adapts the database configuration (e.g. tweaking and cleaning indices~\cite{kephart2003vision}). One difficult and relevant research direction in this area is to create reusable models from machine/deep learning solutions that can be used by several users/companies in different contexts instead of creating multiple solutions from scratch. The bottleneck is that applications/services have peculiarities that may block the direct reuse of solutions for resource optimisation from other users/companies.

Several machine learning and deep learning algorithms require large-scale computing power and external data sources, which can be cheaper and easier to acquire via Cloud than using on-premise infrastructure. This is becoming particularly relevant as technologies to train complex machine/deep learning models can now be executed in parallel at scale \cite{cho2017powerai}. That is why several companies are providing AI-related services in the Cloud such as IBM Watson, Microsoft Azure Machine Learning, AWS Deep Learning AMIs, Google Cloud Machine Learning Engine, among others. Some of these Cloud services can be enhanced while users consume them. This has already delivered considerable savings for CDCs~\cite{gao2014machine}. It can also streamline managed database configuration tuning ~\cite{van2017automatic}.
 
We anticipate a massive adoption of auto-tuners, especially for the SaaS layer of the Cloud. We also foresee the likely advent of new automated tools for Cloud users to benefit from the experience of other users via semi-automated application builders (recommending tools of configurations other similar users have successfully employed), automated database sharders, query optimisers, or smart load balancers and service replicators. As security becomes a key concern for most corporations worldwide, new ML-based security Cloud services will help defend critical Cloud services and rapidly mutate to adapt to new fast-developing threats.

\section{Future Research Directions}
\label{sec:FutureResearch}

The Cloud computing paradigm, like the Web, the Internet, and the computer itself, has transformed the information technology landscape in its first decade of existence. However, the next decade will bring about significant new requirements, from large-scale heterogeneous IoT and sensor networks producing very large data streams to store, manage, and analyse, to energy- and cost-aware personalised computing services that must adapt to a plethora of hardware devices while optimising for multiple criteria including application-level QoS constraints and economic restrictions. 

Significant research was already performed to address the Cloud computing technological and adoption challenges, and the state-of-the-art along with their limitations is discussed thoroughly in Section~\ref{sec:challenges}. The future research in Cloud computing should focus at addressing these limitations along with the problems hurled and opportunities presented by the latest developments in the Cloud horizon. Thus the future R\&D will greatly be influenced/driven by the emerging trends discussed in Section~\ref{sec:trends}. 
Here the manifesto provides the key future directions for the Cloud computing research, for the coming decade. 

\subsection{Scalability and Elasticity}

Scalability and elasticity research challenges for the next decade can be decomposed into hardware, middleware, and application-level.

At the Cloud computing hardware level, an interesting research direction is special-purpose Clouds for specific functions, such as deep learning---e.g. Convolutional Neural Networks (CNNs), Multi-Layer Perceptrons (MLPs), and Long Short-Term Memory (LSTMs)---data stream analytics, and image and video pattern recognition.  While these functionalities may appear to be very narrow, they can be deployed for a spectrum of applications and their usage is increasingly growing. There are numerous examples at control points at airports, social network mining, IoT sensor data analytics, smart transportation, and many other applications. Key Cloud providers are already offering accelerators and special-purpose hardware with increasing usage growth, e.g., Amazon is offering GPUs, Google has been deploying Tensor Processing Units (TPUs)~\cite{jouppi2017datacenter} and Microsoft is deploying FPGAs in the Azure Cloud~\cite{putnam2014reconfigurable}.  As new hardware addresses scalability, Clouds need to embrace non-traditional architectures, such as neuromorphic, quantum computing, adiabatic, nanocomputing and many others (see ~\cite{IEEE:RebootingComputing}). Research needed includes developing appropriate virtualization abstractions, as well as programming abstractions enabling just-in-time compilation and optimisation for special-purpose hardware.  Appropriate economic models also need to be investigated for FaaS Cloud providers (e.g., offering image and video processing as composable micro-services).

At the Cloud computing middleware level, research is required to further increase reuse of existing infrastructure, to improve speed of deployment and provisioning of hardware and networks for very large scale deployments. This includes algorithms and software stacks for reliable execution of applications with failovers to geographically remote private or hybrid Cloud sites.  Research is also needed on InterClouds which will seamlessly enable computations to run on multiple public Cloud providers simultaneously.  In order to support HPC applications, it will be critical to guarantee consistent performance across multiple runs even in the presence of additional Cloud users.  New deployment and scheduling algorithms need to be developed to carefully match HPC applications with those that would not introduce noise in parallel execution or if not possible, to use dedicated clusters for HPC~\cite{gupta2016evaluating,netto2018hpccloud}.  

To be able to address large scale communication-intensive applications, further Cloud provider investments are required to support high throughput and low latency networks \cite{netto2018hpccloud}. The environment of these applications necessitates sophisticated mechanisms for handling multiple clients and for providing sustainable and profitable business provision. Moreover, Big Data applications are leveraging HPC capabilities and IoT, providing support for many modern applications such as smart cities~\cite{santana2016software} or industrial IoT~\cite{bonomi2012fog}. These applications have demanding requirements in terms of (near-)real time processing of large scale of data, its intelligent analysis and then closing the loops of control.


\subsection{Resource Management and Scheduling}
\label{sec:FutureResearch:rms}

The evolution of the Cloud in the upcoming years will lead to a new generation of research solutions for resource management and scheduling. Technology trends such as Fog will increase the level of decentralisation of the computation, leading to increased heterogeneity in the resources and platforms and also to more variability in the processed workloads. Technology trends, such as serverless computing and Edge computing, will also offer novel opportunities to reason on the trade-offs of offloading part of the application logic far from the system core, posing new questions on optimal management and scheduling. Conversely, trends such as software-defined computing and Big Data will come to maturity, expanding the enactment mechanisms and reasoning techniques available for resource management and scheduling, thus offering many outlets for novel research.

Challenges arising from decentralisation are inherently illustrated in the Fog computing domain, edge analytics (discussed further in Section~\ref{sec:FutureResearch:Empowering}) being one interesting research direction. In edge analytics, the stream-based or event-driven sensor data will be processed across the complete hierarchy of Fog topology. This will require cooperative resource management between centralised CDCs and distributed Edge computing resources for real-time processing. Such management methods should be aware of the locations and resources available to edge devices for optimal resource allocation, and should take into account device mobility, highly dynamic network topology, and privacy and security protection constraints at scale. The design of multiple co-existing control loops spanning from CDCs to the Edge is, by itself, a broad research challenge from the point of design, analysis and verification, implementation and testing. The adoption of container technology in these applications will be useful due to its small footprint and fast deployment~\cite{pahl2015containers}.

Novel research challenges in the area of scheduling will also arise in these decentralised and heterogeneous environments. Recently proposed concepts such as multi-resource fairness \cite{ghodsi2011dominant} as well as non-conventional game theoretic methods \cite{semasingheMH17}, which today are primarily applied to small to medium-scale computing clusters or to define optimal economic models for the Cloud, need to be generalised and applied to large-scale heterogeneous settings comprising both CDCs and Edge. For example, mean-field games may help in addressing inherent scalability problems by helping to reason about the interaction of a large number of resources, devices and user types \cite{semasingheMH17}.

Serverless computing is an example of emerging research challenges in management and scheduling, such as offloading the computation far from the application core components that implement the business logic. From the end user standpoint, FaaS raises the expectation that functions will be executed within a specific time, which is challenging given that current performance is quite erratic~\cite{Fox17} and network latency can visibly affect function response time. Moreover, given that function cost is per access, this will require novel resource management policies to decide when and to which extent rely on FaaS instead of microservices that run locally to the application.

From the FaaS provider perspective, allocation of resources needs to be optimal (neither excessive nor insufficient), and, from a user perspective, a desirable level of QoS needs to be achieved when functions are executed, determining suitable trade-offs with execution requirements, network latency, privacy and security requirements. Given that a single application backed by FaaS can lead to hundreds of hits to the Cloud in a second, an important challenge for serverless platform providers will be to optimise allocation of resources for each class of service so that revenue is optimised, while all the user FaaS QoS expectations are met. This research will require to take into consideration soft constraints on execution time of functions and proactive FaaS provisioning to avoid high latency of resource start-up to affect the performance of backed applications. Moreover, providers and consumers, both for FaaS and regular Cloud services, often have different goals and constraints, calling for novel game-theoretic approaches and market-oriented models for resource allocation and regulation of the supply and demand within the Cloud platform.

The emerging SDN paradigm exemplifies a novel trend which will extend the range of control mechanisms available for holistic management of resources. By logically centralising the network control plane, SDNs provide opportunities for more efficient management of resources located in a single administrative domain such as a CDC. SDN also facilitates joint VM and traffic consolidation, a difficult task to do in traditional data centre networks, in order to optimise energy consumption and SLA satisfaction, thus opening new research outlets~\cite{cziva2016sdn}. Service Function Chaining (SFC) is an automated process to set up the chain of virtual network functions (VNFs), e.g.,  network address translation (NAT), firewalls, intrusion detection systems (IDS) in an NFV environment using instantiation of software-only services. Leveraging SDN together with NFV technologies allows for efficient and on-demand placement of service chains~\cite{cho2017real}. However, optimal service chain placement requires novel heuristics and resource management policies. The virtualized nature of VNFs also makes their orchestration and consolidation easier and dynamic deployment of network services possible~\cite{kuo2016deploying,pham2017traffic}, calling for novel algorithms that can exploit these capabilities.

In addition, it is foreseeable that the ongoing interest for ML, deep learning, and AI applications will help in dealing with the complexity, heterogeneity, and scale, in addition to spawn novel research in established data centre resource management problems such as VM provisioning, consolidation, and load balancing. It is however important to recognise that potential loss of control and determinism may arise by adopting these techniques. Research in explainable AI may provide a suitable direction for novel research to facilitate the adoption of AI methods in Cloud management solutions within the industry \cite{doran2017does}.

For example, in scientific workflows the focus so far has been on efficiently managing the execution of platform-agnostic scientific applications. As the amount of data processed increases and extreme-scale workflows begin to emerge, it is important to consider key concerns such as fault tolerance, performance modelling, efficient data management, and efficient resource usage.  For this purpose, Big Data analytics will become a crucial tool~\cite{deelman2017panorama}. For instance, monitoring and analysing resource consumption data may enable workflow management systems to detect performance anomalies and potentially predict failures, leveraging technologies such as serverless computing to manage the execution of complex workflows that are reusable and can be shared across multiple stakeholders. Although today there exist the technical possibility to define solutions of this kind, there is still a shortage of applications of serverless functions to HPC and scientific computing use cases, calling for further research in this space.

\subsection{Reliability}

One of the most challenging areas in Cloud computing systems is reliability as it has a great impact on the QoS as well as on the long term reputation of the service providers. Currently, all the Cloud services are provided based on the cost and performance of the services. The key challenge faced by Cloud service providers is how to deliver a competitive service that meets end users' expectations for performance, reliability, and QoS in the face of various types of independent as well as temporal and spatial correlated failures. So the future of research in this area will be focused on innovative Cloud services that provide reliability and resilience with assured service performance; which is called Reliability as a Service (RaaS). The main challenge is to develop a hierarchical and service-oriented cloud service reliability model based on advanced mathematical and statistical models~\cite{pezoa2012performance}. This requires new modules to be included in the existing Cloud systems such as failure model and workload model to be adapted for resource provisioning policies and provide flexible reliability services to a wide range of applications.  

One of the future directions in RaaS will be using deep and machine learning for failure prediction. This will be based on failure characterisation and development of a model from massive amount of failure datasets. Having a comprehensive failure prediction model will lead to a failure-aware resource provisioning that can guarantee the level of reliability and performance for the user's applications. This concept can be extended as another research direction for the Fog computing where there are several components on the edge. While fault-tolerant techniques such as replication could be a solution in this case, more efficient and intelligent approaches will be required to improve the reliability of new type of applications such as  IoT applications. This needs to be incorporated with the power efficiency of such systems and solving this trade off will be a complex research challenge to tackle~\cite{mehdipour2016fog}.

Another research direction in reliability will be about Cloud storage systems that are now mature enough to handle Big Data applications. However, failures are inevitable in Cloud storage systems as they are composed of large scale hardware components. Improving fault tolerance in Cloud storage systems for Big Data applications is a significant challenge. Replication and Erasure coding are the most important data reliability techniques employed in Cloud storage systems~\cite{nachiappan2017cloud}. Both techniques have their own trade-offs in various parameters such as durability, availability, storage overhead, network bandwidth and traffic, energy consumption and recovery performance. Future research should include the challenges involved in employing both techniques in Cloud storage systems for Big Data applications with respect to the aforementioned parameters~\cite{nachiappan2017cloud}. This hybrid technique applies proactive dynamic data replication of erasure coded data based on node failure prediction, which significantly reduces network traffic and improves the performance of Big Data applications with less storage overhead. So, the main research challenge would be solving a multivariable optimisation problem to take into account several metrics to meet users and providers requirements.

\subsection{Sustainability}
\label{sec:FutureResearch:Sustainability}

Sustainability of ICT systems is emerging as a major consideration~\cite{gelenbe2015impact} due to the energy consumption of ICT systems. Of course, sustainability also covers issues regarding the pollution and decontamination of the manufacturing and decommissioning of computer and network equipment, but this aspect is not covered in the present paper.
 
In response to the concern for sustainability, viewed primarily through the lens of energy consumption and energy awareness, increasingly large CDCs are being established, with up to 1000 MW of potential power consumption, in or close to areas where there are plentiful sources of renewable energy~\cite{berral2014building}, such as hydro-electricity in northern Norway, and where natural cooling can be available as in areas close to the Arctic Circle. This actually requires new and innovative system architectures that can distribute data centres and Cloud computing, geographically. To address this, algorithms have been proposed, which rely on geographically distributed data coordination, resource provisioning and energy-aware and carbon footprint-aware provisioning in data centres~\cite{sustainable-1,sustainable-3,khosravi2017energy}. In addition, geographical load balancing can provide an effective approach for optimising both performance and energy usage. With careful pricing, electricity providers can motivate Cloud service providers to ``follow the renewables'' and serve requests through CDCs located in areas where green energy is available~\cite{liu2015greening}. On the other hand, the smart grid focuses on controlling the flow of energy in the electric grid with the help of computer systems and networks, and there seems to be little if any work on the energy consumption by the ICT components in the smart grid, perhaps because the amount would be small as compared to the overall energy consumption of a country or region. Interestingly enough, there has been recent work on dynamically coupling the flow of energy to computing and communication resources, and the flow of energy to the components of such computer/communication systems \cite{gelenbe2016energy} in order to satisfy QoS and SLAs for jobs while minimising the energy consumption, but much more work will be needed.

However, placing data centres far away from most of the end users places a further burden on the energy consumption and QoS of the networks that connect the end users to the CDCs. Indeed, it is important to note that moving CDCs away from users will increase the energy consumed in networks, so that some remote solutions which are based on renewable energy may substantially increase the energy consumption of networks that are powered through conventional electrical supplies. Another challenge relates to the very short end-to-end delay that certain operations, such as financial transactions, require; thus data centres for financial services often need to be located in proximity to the actual human users and financial organisations (such as banks) that are designing, maintaining and modifying the financial decision making algorithms, as well as to the commodity trading data bases whose state must accurately reflect current prices, since users need to buy and sell stock or other commodities at up-to-date prices that may automatically change within less than a second. Another factor is the proprietary nature of the data that is being used, and the legal and security requirements that can often only be verified and complied within national boundaries or within the EU. Thus if the data remains local, the CDCs that process it also have to be local. Thus in many cases, the Cloud cannot rely on renewable energy to operate effectively simply because renewal energy is not available locally and because some renewable energy sources (e.g. wind and photovoltaic) tend to be intermittent. At the other end, the power needs of CDCs and the Cloud are also growing due to the ever-increasing amount of data that need to be stored and processed. Thus running the Cloud and CDCs in an energy efficient manner remains a major priority.

Unfortunately, high performance and more data processing has always gone hand-in-hand with greater energy consumption. Thus QoS, SLAs, and energy consumption have to be considered simultaneously and need to be managed online~\cite{Lent3}. Since all the fast-changing online behaviours cannot be predicted in advance or modelled in a complete manner, adaptive self-aware techniques are needed to face this challenge~\cite{wang2016adaptive}. Some progress has been recently made in this direction~\cite{wang2018IEEETCC} but further work will be needed. The actual algorithms that may be used will include machine learning techniques such as those described in Yin et al~\cite{yin2017multi}, which exploits constant online measurement of system parameters that can lead to online decision making that will optimise sustainability while respecting QoS considerations and SLAs.

The Fog can also substantially increase energy consumption because of the greater difficulty of efficient energy management for smaller and highly diverse systems~\cite{gelenbe2016energy,Lent2}. At the same time, the reduced access distance and network size from the end users to the Fog servers can create energy savings in networks. Therefore, the interesting trade-off between the increased energy consumption from many disparate and distribute Fog servers, and the reduced network energy consumption when the Fog servers are installed in close proximity to the end user, requires much further work \cite{Lent1}. Such research should include the improvements in network QoS that may be experienced by end users, when they access locally distributed Fog servers and their traffic traverses a smaller number of network nodes. There have been attempts to conduct experimental research in this direction with the help of machine learning based techniques \cite{wang2016adaptive}.
 
Some approaches for improving sustainability and reducing energy consumption in the Cloud, primarily focus on the VM consolidation for minimising the energy consumption of the servers, which has been shown to be quite effective~\cite{berl2010energy}, while the Cloud cannot be accessed without the help of networks. However, reducing energy consumption in networks is also a complex problem~\cite{Morfo,Francois1}. Saving energy for networking elements often disturbs other aspects such as reliability, scalability, and performance of the network~\cite{Silvestri}. Proposals have been made and tested regarding the design of smart energy-aware routing algorithms~\cite{Mahmoodi}, but this area in general has received less attention compared to energy consumption and power efficiency of computing elements. With the advent of SDN, the global network awareness and centralised decision-making offered by SDN may provide a better opportunity for creating sustainable networks for Clouds~\cite{Francois2}. This is perhaps one of the areas that will draw substantially more research efforts and innovation in the next decade.

\subsection{Heterogeneity}

Heterogeneity on the Cloud was introduced in the last decade, but awaits widespread adoption. As highlighted in Section~\ref{subsec:SOA:Heterogeneity}, there are currently at least two significant gaps that hinder heterogeneity from being fully exploited on the Cloud. The first gap is between unified management platforms and heterogeneity. Existing research that targets resource and workload management in heterogeneous Cloud environments is fragmented. This translates into the lack of availability of a unified environment for efficiently exploiting VM level, vendor level and hardware architecture level heterogeneity while executing Cloud applications. The manifesto therefore proposes for the next decade an umbrella platform that accounts for heterogeneity at all three levels. This can be achieved by integrating a portfolio of workload and resource management techniques from which optimal strategies are selected based on the requirement of an application. For this, heterogeneous memory management will be required. Current solutions for memory management rely mainly on hypervisors, which limits the benefits from heterogeneity. Alternate solutions recently proposed rely on making guest operating systems heterogeneity-aware~\cite{heterodirection-1}.

The second gap is between abstraction and heterogeneity. Current programming models for using hardware accelerators require accelerator specific languages and low level programming efforts. Moreover, these models are conducive for developing scientific applications. This restricts the wider adoption of heterogeneity for service oriented and user-driven applications on the Cloud. One meaningful direction to pursue will be to initiate a community-wide effort for developing an open-source high-level programming language that can satisfy core Cloud principles, such as abstraction and elasticity, which are suited for modern and innovative Cloud applications in a heterogeneous environment. This will also be a useful tool as the Fog ecosystem emerges and applications migrate to incorporate both Cloud and Fog resources. 

Recent research in this area has highlighted the limitation of current programming languages, such as OpenCL~\cite{heterodirection-2}. The interaction between CPUs and the hardware accelerator need to be explicitly programmed, which limits the automatic transformation of source code in efficient ways. To this end, fine-grained task partitioning needs to be automated for general purpose applications. Additionally, the automated conversion from coarse-grained to fine-grained task partitioning is required.  In the context of OpenCL programming, there is limited performance portability, which is to be addressed. However, currently available high-level programming languages, such as TANGRAM~\cite{heterodirection-3} provide performance portability across different accelerators, but need to incorporate performance models and adaptive runtimes for finding optimal strategies for interaction between the CPU and the hardware accelerator.

Although the Cloud as a utility is a more recent offering, a number of the underlying technologies for supporting different levels of heterogeneity (memory, processors etc) in the Cloud came into inception a few decades ago. For example, the Multiplexed Information and Computing Service (Multics) offered single-level memory, which was the foundation of virtual memory for heterogeneous systems. Similarly, IBM developed CP-67, which was one of the first attempts in virtualizing mainframe operating systems to implement time-sharing. Later on VMWare used this technology for virtualizing x86 servers. The earlier technology was able to even provide I/O virtualization, and meaningful ways of addressing some of the challenges raised by modern heterogeneity may find inspiration in earlier technologies when the Cloud was not known.

Recently there is also a significant discussion about disaggregated data centres. Traditionally data centres are built using servers and racks with each server contributing the resources such as CPU, memory and storage, required for the computational tasks. With the disaggregated data centre each of these resources is built as a stand-alone resource ``{\em blade}'', where these blades are interconnected through a high-speed network fabric. The trend has come into existence as there is significant gap in the pace at which each of these resource technologies individually advanced.
Even though most prototypes are proprietary and in their early stages of development, a successful deployment at the data centre level would have significant impact on the way the traditional IaaS are provided. However, this needs significant development in the network fabric as well~\cite{gao2016network}.

\subsection{Interconnected Clouds}

As the grid computing and web service histories have shown, interoperability and portability across Cloud systems is a highly complicated area and it is clear at this time that pure standardisation is not sufficient to address this problem. The use of application containers and configuration management tools for portability, and the use of software adapters and libraries for interoperability are widely used as practical methods for achieving interoperation across Cloud services and products. However, there are a number of challenges~\cite{buyya2010intercloud}, and thus potential research directions, that have been around since the early days of Cloud computing and, due to their complexity, have not been satisfactorily addressed so far.

One of such challenges is how to promote Cloud interconnection without forcing the adoption of the minimum common set of functionalities among services: if users want, they should be able to integrate complex functionalities even if they are offered only by one provider. Other research directions include how to enable Cloud interoperation middleware that can mimic complex services offered by one provider by composing simple services offered by one or more providers - so that the choice about the complex service or the composition of simpler services were solely dependent on the user constraints - cost, response time, data sovereignty, etc.

The above raises another important future research direction: how to enable middleware operating at the user-level (InterCloud and hybrid Clouds) to identify candidate services for a composition without support from Cloud providers? Given that providers have economic motivation to try to retain all the functionalities offered to their customers (i.e., they do not have motivation to facilitate that only some of the services in a composition are their own), one cannot expect that an approach that requires Cloud providers cooperation might succeed. 

Therefore, the middleware enabling composition of services has to solve challenges in its two interfaces: in the interface with Cloud users, it needs to seamlessly deliver the service, in a level where how the functionality is delivered is not relevant for users: it could be obtained in all from a single provider (perhaps invoking a SaaS able to provide the functionality) or it could be obtained by composing different services from different providers. In the provider interface, it enables such more complex functions to be obtained, regardless of particular collaboration from providers: provided that an API exists, the middleware would be in charge of understanding what information/service the API can provide (and how to access such service) and thus decide by itself if it has all the required input necessary to access the API, and even the output is sufficient to enable the composition. This discussion makes clear the complexity of such middleware and the difficulty of the questions that need to be addressed to enable such vision.

Nevertheless, ubiquitously interconnected Clouds (achieved via Cloud Federation) can truly be achieved only when Cloud vendors are convinced that the Cloud interoperability adoption brings them financial and economic benefits. This requires novel approaches for billing and accounting, novel interconnected Cloud suitable pricing methods, along with formation of InterCloud marketplaces~\cite{toosi2014interconnected}.


Finally, the emergence of SDNs and the capability to shape and optimise network traffic has the potential to influence research in Cloud interoperation. Google reports that one of the first uses of SDNs in the company was for optimisation of wide-area network traffic connecting their data centres~\cite{vahdat2015purpose}. In the same direction, investigation is needed on the feasibility and benefits of SDN and NFV to address some of the challenges above. For example, SDN and NFV can enable better security and QoS for services built as compositions of services from multiple providers (or from geographically distributed services from the same provider) by enforcing prioritization of service traffic across providers/data centres and specific security requirements~\cite{huedo2017interoperable}.

\subsection{Empowering Resource-Constrained Devices}
\label{sec:FutureResearch:Empowering}

Regarding future directions for empowering resource-constrained devices, in the mobile Cloud domain, we already have identified that, while task delegation is a reality, code offloading still has adaptability issues. It is also observed that, \textit{``as the device capabilities are increasing, the applications that can benefit from the code offloading are becoming limited''}~\cite{srirama2017mobile}. This is evident, as the capabilities of smartphones are increasing, to match or benefit from offloading, the applications are to be offloaded to Cloud instances with much higher capacity. This incurs higher cost per offloading. To address this, the future research in this domain should focus at better models for multi-tenancy in Mobile Cloud applications, to share the costs among multiple mobile users. The problem further gets complex due to the heterogeneity of both the mobile devices and Cloud resources.

We also foresee the need for incentive mechanisms for heterogeneous mobile Cloud offloading to encourage mobile users to participate and get appropriate rewards in return. This should encourage in adapting the mobile Cloud pattern to the social networking domain as well, in designing ideal scenarios. In addition, the scope and benefits offered by the emerging technologies such as serverless computing, CaaS and Fog computing, to the mobile Cloud domain, are not yet fully explored.
 
The incentive mechanisms are also relevant for the IoT and Fog domains. Recently there is significant discussion about the establishment of Fog closer to the \textit{things}, by infrastructure offered by independent Fog providers~\cite{ChangETAL:indieFog.17}. These architectures follow the consumer-as-provider (CaP) model. A relevant CaP example in the Cloud computing domain is the MQL5 Cloud Network~\cite{MQL5.17}, which utilises consumer's devices and desktops for performing various distributed computing tasks. Adaptation of such Peer-to-Peer (P2P) and CaP models would require ideal incentive mechanisms. Further discussion about the economic models for such Micro Data centres is provided in Section~\ref{subsec:future:economics}.

The container technology also brings several opportunities to this challenge. With the rise of Fog and Edge computing, it can be predicted that the container technology, as a kind of lightweight running environment and convenient packing tools for applications, will be widely deployed in edge servers. For example, the customised containers, such as Cloud Android Container~\cite{wu2017container}, aimed at Edge computing and offloading features will be more and more popular. They provide efficient server runtime and inspire innovative applications in IoT, AI, and other promising fields.
 
Edge analytics in domains such as real-time streaming data analytics would be another interesting research direction for the resource constrained devices. The things in IoT primarily deal with sensor data and the Cloud-centric IoT (CIoT) model extracts this data and pushes it to the Cloud for processing. Primarily, Fog/Edge computing came to existence in order to reduce the network latencies in this model. In edge analytics, the sensor data will be processed across the complete hierarchy of Fog topology, i.e. at the edge devices, intermediate Fog nodes and Cloud. The intermediary processing tasks include filtering, consolidation, error detection etc. Frameworks that support edge analytics (e.g. Apache Edgent~\cite{Edgent.18}) should be studied considering both the QoS and QoE (Quality of Experience) aspects. Preliminary solutions related to scheduling and placement of the edge analytics tasks and applications across the Fog topology are already appearing in the literature~\cite{soo2017proactive,mahmudquality}. Further research is required to deal with cost-effective multi-layer Fog deployment for multi-stage data analytics and dataflow applications.

\subsection{Security and Privacy}

Security and privacy issues are among the biggest concerns in adopting Cloud technologies. In particular, security and privacy issues are related to various technologies including, networks (Section~\ref{sec:FutureResearch:Networking}), databases, virtualization, resource scheduling (Section~\ref{sec:FutureResearch:rms}), and so on. Possible solutions must be designed according to the specific trust assumptions at the basis of the considered scenario (e.g., a Cloud provider can be assumed completely untrusted/malicious, or it could be assumed trustworthy). In the following, we provide a brief description of future research directions in the security and privacy area, mainly focusing on problems related to the management of (sensitive) data. 

Regarding the protection of data in the Cloud, we distinguish between two main scenarios of future research: 1) a simple scenario where the main problem is to guarantee the protection of data in storage as well as the ability to efficiently access and operate on them; 2) a scenario where data must be shared and accessed by multiple users and with the possible presence of multiple providers for better functionality and security. In the simple scenario, when data are protected with client-side encryption, there is the strong need for scalable and well-performing techniques that, while not affecting service functionality, can: 1) be easily integrated with current Cloud technology; 2) avoid possible information leakage caused by the solutions (e.g., indexes) adopted for selectively retrieving data or by the encryption supporting queries~\cite{naveed2015inference}; 3) support a rich variety of queries. Other challenges are related to the design of solutions completely departing from encryption and based on the splitting of data among multiple providers to guarantee generic confidentiality and access/visibility constraints possibly defined by the users.  Considering the data integrity problem, an interesting research direction consists in designing solutions proving data integrity when data are distributed and stored on multiple independent Cloud providers. In the scenario with multiple users and possible multiple providers, a first issue to address is the design of solutions for selectively sharing data that support: 1) write privileges as well as multiple writers; 2) the efficient enforcement of policies updates in distributed storage systems characterised by multiple and independent Cloud providers; 3) the selective sharing of information among parties involved in distributed computations, thus also taking advantage of the availability of cheaper (but not completely trusted) Cloud providers. The execution of distributed computations also requires the investigation of issues related to query privacy (which deals with the problem of protecting accesses to data) and computation integrity. Existing solutions for query privacy are difficult to apply in real-world scenarios for their computational complexity or for the limited kinds of queries supported. Interesting open issues are therefore the development of scalable and efficient techniques: i) supporting concurrent accesses by different users; and ii) ensuring no improper leakage on user activity and applicability in real database contexts. With respect to computation integrity, existing solutions are limited in their applicability,  the integrity guarantees offered, and  the kinds of supported queries. There is then the need to design a generic framework for evaluating the integrity guarantees provided according to the cost that a user is willing to pay to have such guarantees and that support different kinds of queries/computations.  In presence of multiple Cloud providers offering similar services, it is critical for users to select the provider that better fits their need.  Existing solutions supporting users in this selection process consider only limited user-based requirements (e.g., cost and performance requirements only) or pre-defined indicators. An interesting challenge is therefore the definition of a comprehensive framework that allows users both to express different requirements and preferences for the Cloud provider selection, and to verify that Cloud providers offer services fully compliant with the signed contract.

While emerging scenarios such as Fog Computing (Section~\ref{ss:bigdata}) and Big Data (Section~\ref{subsec:fog}) have brought enormous benefits, as a side effect there is a tremendous exposure of private and sensitive information to privacy breaches. The lack of central controls in Fog-based scenarios may raise privacy and trust issues. Also, Fog computing assumes the presence of trusted nodes together with malicious ones. This requires adapting the earlier research of secure routing, redundant routing and trust topologies performed in the P2P context, to this novel setting~\cite{garcia2015edge}. While Cloud security research can rely on the idea that all data could be dumped into a data lake and analysed (in near real time) to spot security and privacy problems, this may no longer be possible when devices are not always connected and there are too many of them to make it financially viable to dump all the events into a central location. This Fog-induced fragmentation of information combined with encryption will foster a new wave of Cloud security research.  Also the explosion of data and their variety (i.e., structured, unstructured, and semi-structured formats) make the definition and enforcement of scalable data protection solutions a challenging issue, especially considering the fact that the risk of inferring sensitive information significantly increases in Big Data. Other issues are related to the provenance and quality of Big Data. In fact, tracking Big Data provenance can be useful for: i) verifying whether data came from trusted sources and have been generated and used appropriately; and ii) evaluating the quality of the Big Data, which is particularly important in specific domains (e.g., healthcare). Blockchain technology can be helpful for addressing the data provenance challenge since it ensures that data in a blockchain are immutable, verifiable, and traceable. However, it also introduces novel privacy concerns since data (including personal data) in a blockchain cannot be changed or deleted.

At the infrastructure level, security and privacy issues that need to be further investigated include: the correct management of virtualization enabling multi-tenancy in the Cloud; the allocation and de-allocation of resources associated with virtual machines as well as the placement of virtual machine instances in the Cloud in accordance to security constraints imposed by users; and the identification of legitimate request to tackle issues such as Denial of Service (DoS) or other forms of cyber-attacks. These types of attacks are critical, as a coordinated attack on the Cloud services can be wrongly inferred as legitimate traffic and the resources would be scaled up to handle them. This will result in both the incurred additional costs and waste in energy~\cite{somani2017ddos}. Cloud systems should be able to distinguish these attacks and decide either to drop the additional load or avoid excessive provisioning of resources. This requires extending the existing techniques of DDoS to also include exclusive characteristics of Cloud systems.

\subsection{Economics of Cloud Computing}
\label{subsec:future:economics}

The economics of Cloud computing offers several interesting future research directions. As Cloud computing deployments based on VMs transition to the use of container-based deployments, there is increasing realisation that the lower overheads associated with container deployment can be used to support real-time workloads. Hence, serverless computing capability is now becoming commonplace with Google Cloud Functions, Amazon Lambda, Microsoft Azure Functions and IBM Bluemix OpenWhisk. In these approaches, no computing resources are actually charged for until a function is called. These functions are often simpler in scope and typically aimed at processing data stream-based workloads. The actual benefit of using serverless computing depends on the execution behaviour and types of workloads expected within an application. Eivy~\cite{eivy2017wary} outlines the factors that influence the economics of such function deployment, such as: (1) average vs. peak transaction rates; (2) scaling number of concurrent activity on the system, i.e. running multiple concurrent functions with increasing number of users; (3) benchmark execution of serverless functions on different backend hardware platforms, and the overall execution time required for your function. 

Similarly, increasing usage of Fog and Edge computing capabilities alongside Cloud-based data centres offers significant research scope in Cloud economics. The combination of stable Cloud resources and volatile user edge resources can reduce the operating costs of Cloud services and infrastructures. However, we expect users to require some incentives to make their devices available at the edge.  The availability of Fog and Edge resources provides the possibility for a number of additional business models and the inclusion of additional category of providers in the Cloud marketplace. We refer to the existence of such systems as Micro Data Centres (MDCs), which are placed between the more traditional data centre and user owned/provisioned resources. Business models include:  
(1) \textit{Dynamic MDC discovery:} in this model, a user would dynamically be able to choose a MDC provider, according to the MDC availability profile, security credentials, or type. A service-based ecosystem with multiple such MDC providers may be realised, however this will not directly guarantee the fulfilment of the user objectives through integration of externally provisioned services. (2) \textit{Pre-agreed MDC contracts:} in this model, detailed contracts adequately capture the circumstances and criteria that influence the performance of the MDC provisioned external services. A user's device would have these pre-agreed contracts or SLA with specific MDC operators, and would interact with them preferentially. This also reduces the potential risks incurred by the user. In performance-based contracts, an MDC would need to provide a minimum level of performance (e.g. availability) to the user which is reflected in the associated price. This could be achieved by interaction between MDCs being managed by the same operator, or by MDC outsourcing some of their tasks to a CDC; (3) \textit{MDC federation:} in this model multiple MDC operators can collaborate to share workload within a particular area, and have preferred costs for exchange of such workload. This is equivalent to alliances established between airline operators to serve particular routes. To support such federation, security credentials between MDCs must be pre-agreed. This is equivalent to an extension of the pre-agreed MDC contracts business model, where MDCs across multiple coffee shop chains can be federated, offering greater potential choice for a user; (4) \textit{MDC-Cloud data centre exchange:} in this model a user's device would contact a CDC in the first instance, which could then outsource computation to an MDC if it is unable to meet the required QoS targets (e.g. latency). A CDC could use any of the three approaches outlined above i.e. dynamic MDC discovery, preferred MDCs, or choice of an MDC within a particular group. A CDC operator needs to consider whether outsourcing could still be profitable given the type of workload a user device is generating.

However, the unpredictable Cloud environment arising due to the use of Fog and Edge resources, and the dynamics of service provisioning in these environments, requires architects to embrace uncertainty. More specifically, architecting for the Cloud needs to strike a reasonable balance between dependable and efficient provision and their economics under uncertainties. In this context, the architecting process needs to incubate architecture design decisions that not only meet qualities such as performance, availability, reliability, security, compliance, among others, but also seek value through their provision. Research shall look at possible abstractions and formulations of the problem, where competitive and/or cooperative game design strategies can be explored to dynamically manage various players, including Cloud multitenants, service providers, resources etc. Future research should also explore Cloud architectures and market models that embrace uncertainties and provide continuous ``win-win'' resolutions (for providers, users and intermediaries) for value and dependability.

Similarly, migrating in-house IT systems (e.g. Microsoft Office 365 for managing email) and IT departments (e.g. systems management) to the Cloud also offers several research opportunities. What this migration means, longer term, for risk tolerance and business continuity remains unclear. Many argue that outsourcing of this kind gives companies access to greater levels of expertise (especially in cybersecurity, software updates, systems availability, etc.) compared to in-house management. However, issues around trust remain for many users -- i.e. who can access their data and for what purpose. Recent regulations, such as the European GDPR and US CLOUD Act are aimed at addressing some of these concerns. The actual benefit of GDPR will probably not be known for a few years, as it comes into effect towards the end of May 2018.

The Edge analytics discussed in Section~\ref{sec:FutureResearch:Empowering} also offers several research directions in this regard. Understanding what data should remain at or near user premises, and what should be migrated for analysis at a data centre remain important challenges. These also influence potential revenue models that could be developed taking account of a number of potential data storage/processing actors that would now exist from the data capture site to subsequent analysis within a CDC. 

In addition, the Cloud Market place today is continuously expanding, with Cloud Harmony provider directory~\cite{CloudHarmony:18} reporting over 90 Cloud providers today. Although some providers dominate the market, there is still significant potential for new players to emerge, especially with recent emphasis on edge and serverless computing. Edge computing, in particular, opens up the potential market to telco operators who manage the mobile phone infrastructure. With increasing data volumes from emerging application areas such as autonomous vehicles and smart city sensing, such telco vendors are likely to form alliances with existing Cloud providers for supporting real time stream processing and edge analytics.

\subsection{Application Development and Delivery}
\label{subsec:future:appdev}

Agile, continuous, delivery paradigms often come at the expense of reduced reasoning at design-time on quality aspects such as SLA compliance, business alignment, and value-driven design, posing for example a risk of adopting the wrong architecture in the early design stages of a new Cloud application. These risks raise many research challenges on how to continuously monitor and iteratively evolve the design and quality of Cloud applications within the continuous delivery pipelines. The definition of supporting methods, high-level programming abstractions, tools and organisational processes to address these challenges is currently a limiting factor that requires further research. For example, it is important to extend existing software development and delivery methodologies with reusable abstractions for designing, orchestrating and managing IoT, Fog/Edge computing, Big Data, and serverless computing technologies and platforms. Early efforts in these directions are already underway~\cite{casale2016current}. 

The trend towards using continuous delivery tools to automatically create, configure, and manage Cloud infrastructures (e.g., Chef, Ansible, Puppet, etc) through infrastructure-as-code is expected to continue and grow in the future years. However, there is still a fundamental shortage of software engineering methods specifically tailored to write, debug and evolve infrastructure-as-code. A challenge here is that infrastructure-as-code is often written in a combination of different programming and scripting languages, requiring greater generality than today in designing software quality engineering tools.

Another direction to extend existing approaches to Cloud application development and delivery is to define new architectural styles and Cloud-native design patterns to make Cloud application definition a process closer to human-thinking than today. The resulting software architectures and patterns need to take into account the runtime domain, and tolerate changes in contexts, situations, technologies, or service-level agreements leveraging the fact that, compared to traditional web services, emerging microservices and architectures offer simpler ways to automatically scale capacity, parallelism, and large-scale distribution, e.g., through microservices, serverless and FaaS. 

Among the main challenges, the definition of novel architectures and patterns needs in particular to tackle Cloud application decomposition. The rapid growth of microservices and the fact that containers are becoming a de facto standard, raises the possibility to decompose an application in many more ways than in the past, with implications on its security, performance, reliability, and operational costs.

Further to this, with serverless computing and FaaS there will be the need for developing novel integration and control patterns to define services that combine traditional external services along with the serverless computing services. As an example, bridging in Edge computing the gap between cyber-physical systems (sensors, actuators, control layer) and the Cloud requires patterns to assist developers in building Cloudlets/swarmlets~\cite{lee2014swarm}. These are fragments of an application making local decisions and delegating tasks that cannot be solved locally to other Cloudlets/swarmlets in the Cloud~\cite{flores2014mobile}, which are further discussed in Section~\ref{subsec:SOA:Empowering}. Developing effective Cloud design patterns also requires fundamental research on meta-controls for dynamic and seamless switching between these patterns at runtime, based on their value potentials and prospects. Such meta-controllers may rely on software models created by the application designers. Proposals in this direction include model-driven engines to facilitate reasoning, what-if analysis, monitoring feedback analysis, and for the correct enactment of adaptation decisions~\cite{BergmayrBFRSWKL18, nitto2017model}. 

Further research in patterns and architectures that combine multiple paradigms and technologies, will also require more work on formalisms to describe the user workload. Requirements in terms of performance, reliability, and security, need to be decomposed and propagated in architectures that combine emerging technologies (e.g., blockchain, SDN, Spark, Storm etc.) giving the ability not just to express execution requirements, but also to characterise the properties of the data processed by the application.

The trade-offs of orchestration of such integrated service mixes need to be investigated systematically considering the influence of the underpinning choice of Cloud resources (e.g., on-demand, reserved, spot, burstable) and the trade-off arising across multiple quality dimensions: (i) security (e.g., individual functions are easier to protect and verify than monoliths vs. greater attack surface with FaaS-based architectures); (ii) privacy (e.g., the benefits of model-based orchestration of access control vs greater data exposure in FaaS because of function calls and data flows); (iii) performance (e.g., the benefits of function-level autoscaling vs increased network traffic and latency experienced with FaaS); (iv) cost (e.g., FaaS cheaper to use per function invocation but can incur higher network charges than other architectural styles).

Research is also needed in programming models for adaptive elastic mobile decentralised distributed applications as needed by Fog/Edge computing, InterClouds, and the IoT.  Separation of concerns will be important to address complexity of software development and delivery models.  Functional application aspects should be specified, programmed, tested, and verified modularly.  Program specifications may be probabilistic in nature, e.g., when analysing asynchronous data streams.  Research is needed in specifying and verifying correctness of non-deterministic programs, which may result, e.g., from online machine learning algorithms.  Non-functional aspects, e.g., fault tolerance, should be translucent:  they can be completely left to the middleware, or applications should have declarative control over them, e.g., a policy favouring execution away from a mobile device in battery-challenged conditions~\cite{beloglazov2012energy}. {\it Translucent} programming models, languages, and Application Programming Interfaces (APIs) will be needed to enable tackling the complexity of application development while permitting control of application delivery to future-generation Clouds.
One research direction to pursue will be the use of even finer-grained programming abstractions such as the actor model and associated middleware to dynamically reconfigure programs between edge resources and CDCs through transparent migration for users~\cite{imai2012elastic,varela-pdcs-mitpress-2013}.

\subsection{Data Management}


While Cloud IaaS and PaaS service for storage and data management focus on file, semi-structured and structured data independently, there is not much explicit focus on metadata management for datasets. Unlike structured data warehouses, the concept of ``Data Lakes'' encourages enterprises to put all their data into Cloud storage, such as HDFS, to allow knowledge to be mined from it. However, a lack of tracking metadata describing the source and provenance of the data makes it challenging to use them. Scientific repositories have over a decade of experience with managing large and diverse datasets along with the metadata that gives a context of use. Provenance that tracks the processing steps that have taken place to derive a data is also essential, for data quality, auditing and corporate governance. S3 offers some basic versioning capability, but metadata and provenance do not yet form a first-class entity in Cloud data platforms.

A key benefit of CDCs is the centralised collocation and management of data and compute at globally distributed data centres, offering economies of scale. The latency to access to data is however a challenge, along with bandwidth limitations across global networks. While Content Distribution Networks (CDN) such as AWS CloudFront cache data at regional level for web and video delivery, these are designed for slow-changing data and there is no such mechanism to write in data closer to the edge. Having Cloud data services at the Fog layer, which is a generalisation of CDN is essential. This is particularly a concern as IoT and 5G mobile networks become widespread.

In addition, Cloud storage has adapted to emerging security and privacy needs with support for HIPAA (Health Insurance Portability and Accountability Act of 1996) and other US CLOUD Act and EU GDPR regulations for data protection. However, enterprises that handle data that is proprietary and have sensitive trade secrets that can be compromised, if it is accessed by the Cloud provider, still remains a concern. While legal protections exist, there are no clear audit mechanisms to show that data has not been accessed by the Cloud provider themselves. Hybrid solutions where private data centres that are located near the public CDCs with dedicated high-bandwidth network allow users to manage sensitive data under their supervision while also leveraging the benefits of public Clouds~\cite{NetApp.18}.

Similarly, the interplay between hybrid models and SDN as well as joint optimisation of data flow placement, elasticity of Fog computing and flow routing can be better explored. Moreover, the computing capabilities of network devices can be leveraged to perform in-transit processing. The optimal placement of data processing applications and adaptation of dataflows, however, are hard problems. This problem becomes even more challenging when considering the placement of stream processing tasks along with allocating bandwidth to meet latency requirements. 

Furthermore, frameworks that provide high-level programming abstractions, such as Apache Bean, have been introduced in recent past to ease the development and deployment of Big Data applications that use hybrid models. Platform bindings have been provided to deploy applications developed using these abstractions on the infrastructure provided by commercial public Cloud providers such as Google Cloud Engine, Amazon Web Services, and open source solutions. Although such solutions are often restricted to a single cluster or data centre, efforts have been made to leverage resources from the edges of the Internet to perform distributed queries or to push frequently-performed analytics tasks to edge resources. This requires providing means to place data processing tasks in such environments while minimising the network resource usage and latency. In addition, efficient methods are to be investigated which manage resource elasticity in such scenarios. Moreover, high-level programming abstractions and bindings to platforms capable of deploying and managing resources under such highly distributed scenarios are desirable.

Lastly, there is a need to examine specialised data management services to support the trifecta of emerging disruptive technologies that will be hosted on Clouds: Internet of Things, Deep Learning, and Blockchain. As mentioned above, IoT will involve a heightened need to deal with streaming data, their efficient storage and the need to seamlessly incorporate data management on the edge seamlessly with management in the Cloud. Trust and provenance is particularly important when unmanaged edge devices play an active role.

The growing role of deep learning (see Section \ref{sec:deeplearning}) will place importance on efficient management of trained models and their rapid loading and switching to support online and distributed analytics applications. Training of the models also requires access to large datasets, and this is particularly punitive for video and image datasets that are critical for applications like autonomous vehicles, and augmented reality. Novel data management techniques that offer compact storage and are also aware of the access patterns during training will be beneficial.

Lastly, Blockchain and distributed ledgers (see Section~\ref{sec:blockchain}) can transform the way we manage and track data with increased assurance and provenance~\cite{mckinsey:blockchain-data}. Financial companies (with crypto-currencies being just a popular manifestation) are at the forefront of using them for storing and tracking transactions, but these can also be extended to store other enterprise data in a secure manner with an implicit audit trail. Cloud-hosted distributed ledgers are already available as generic implementations (e.g. Ethereum and Hyperledger Fabric Blockchain platforms) but these are likely to be incorporated as integral part of Cloud data management. Another interesting area of research is in managing the ledger data itself in an efficient and scalable manner.

\subsection{Networking}
\label{sec:FutureResearch:Networking}

Global network view, programmability, and openness features of SDN provide a promising direction for application of SDN-based traffic engineering mechanisms within and across CDC networks. By using SDN within a data centre network, traffic engineering (TE) can be done much more efficiently and intelligently with dynamic flow scheduling and management based on current network utilisation and flow sizes~\cite{al2010hedera}.  Even though traffic engineering has been widely used in data networks, distinct features of SDN need a novel set of traffic engineering methods to utilise the available global view of the network and flow characteristics or patterns~\cite{akyildiz2014roadmap}. During the next decade we will also expect to see techniques targeting network performance requirements such as delay and bandwidth or even jitter guarantees to comply with QoS requirements of the Cloud user application and enforce committed SLAs.

SDN may also influence the security and privacy challenges in Cloud. In general, within the networking community, the overall perception is that SDN will help improve security and reliability both within the network-layer and application-layer. As suggested by Kreutz et al~\cite{kreutz2015software}, the capabilities brought by SDN may be used to introduce novel security services and address some of the on-going issues in Clouds. These include but are not limited to areas such as policy enforcement (for example, firewalling, access control, middleboxes), DoS attack detection and mitigation, monitoring infrastructures for fine-grained security examinations, and traffic anomaly detection.

Nevertheless, as a new technology, the paradigm shift brought by SDN brings along new threat vectors that may be used to target the network itself, services deployed on SDNs and the associated users. For instance, attackers may target the SDN controller as the single point of attack or the inter-SDN communications between the control and data plane - threats that did not exist in traditional networks. At the same time, the impact of existing threats may be magnified such as the range of capabilities available to an adversary who has compromised the network forwarding devices~\cite{shaghaghi2017wedgetail}. Hence, importing SDN to Clouds may impact the security of Cloud services in ways that have not been experienced or expected, which requires further research in this area. 

The Cloud community has given significant priority to intra data centre networking, while efficient solutions for networking in interconnected environments are also highly demanded. Recent advances in SDN technology are expected to simplify intra data centre networking by making networks programmable and reduce both capital and operational expenditure for Cloud providers. However, the effectiveness of current approaches for interconnected Cloud environments and how SDN is used over public communication channels need further investigation.

One of the areas of networking that requires more attention is the management and orchestration of NFV environments. SFC is also a hot topic attaining a significant amount of attention by the community. So far, little attention has been paid to virtual network function (VNF) placement and consolidation while meeting the QoS requirements of the applications is highly desirable. Auto-scaling of VNFs within the service chains also requires in-depth attention. VNFs providing networking functions for the applications are subject to performance variation due to different factors such as the load of the service or overloaded underlying hosts. Therefore, development of auto-scaling mechanisms that monitor the performance of VNF instances and adaptively add or remove VNF instances to satisfy the SLA requirements of the applications is of paramount importance. Traffic engineering combined with migration and placement of VNFs provide a promising direction for the minimisation of network communication cost. Moreover, in auto-scaling techniques, the focus is often on auto-scaling of a single network service (e.g., firewall), while in practice auto-scaling of VNFs must be performed in accordance with service chains.

Recent advances in AI, ML, and Big Data analytics have great potential to address networking challenges of Cloud computing and automation of the next-generation networks in Clouds. The potential of these approaches along with centralised network visibility and readily accessible network information (e.g., network topology and traffic statistics) that SDN brings into picture, open up new opportunities to use ML and AI in networking. Even though it is still unclear how these can be incorporated into networking projects, we expect to see this as one of the exotic research areas in the following decade.  

The emergence of IoT connecting billions of devices all generating data will place major demands on network infrastructure. 5G wireless and its bandwidth increase will also force significant expansion in network capacity with explosion in the number of mobile devices. Even though a key strategy in addressing latency and lower network resource usage is Edge/Fog computing, Edge/Fog computing itself is not enough to address all the networking demand. To meet the needs of this transition, new products and technologies expanding bandwidth, or the carrying capacity, of networks are required along with advances in faster broadband technologies and optical networking. Moreover, in both Edge and Fog computing, the integration of 5G so far has been discussed within a very narrow scope. Although 5G network resource management and resource discovery in Edge/Fog computing have been investigated, many other challenging issues such as topology-aware application placement, dynamic fault detection, and network slicing management in this area are still unexplored.

\subsection{Usability}

There are several opportunities to enhance usability in Cloud environments. For instance, it is still hard for users to know how much they will spend renting resources due to workload/resource fluctuations or characteristics. Tools to have better estimations would definitely improve user experience and satisfaction. Due to recent demands from Big Data community, new visualization technologies could be further explored on the different layers of Cloud environment to better understand infrastructure and application behaviour and highlight insights to end users. Easier API management methodologies, tools, and standards are also necessary to handle users with different levels of expertise and interests. User experience when handling data-intensive applications also needs further studies considering their expected QoS. 

In addition, users are still overloaded with resource and service types available to run their applications. Examples of resources and services are CPUs, GPUs, network, storage, operating system flavour, and all services available in the PaaS. Advisory systems to help these users would greatly enhance their experience consuming Cloud resources and services. Advisory systems to also recommend how users should use Cloud more efficiently would certainly be beneficial. Advices such as whether data should be transferred or visualized remotely, whether resources should be allocated or deleted, whether baremetal machines should replace virtual ones are examples of hints users could receive to make Cloud easier to use and more cost-effective.

The main difficulty in this area lies on evaluation. Traditionally, Cloud computing researchers and practitioners mostly perform quantitative experiments, whereas researchers working closer to users have deep knowledge on qualitative experiments. This second type of experiments depends on selecting groups of users with different profiles and investigating how they use technology. As Cloud has a very heterogeneous community of users with different needs and skills and work in different Cloud layers (IaaS, PaaS, and SaaS), such experiments are not trivial to be designed and executed at scale. Apart from understanding user behaviour, it is relevant to develop mechanisms to facilitate or automatically reconfigure Cloud technologies to adapt to the user needs and preferences, and not assume all users have the same needs or have the same level of skills.

\subsection{Discussion}

\begin{figure}
  \includegraphics[width=\textwidth]{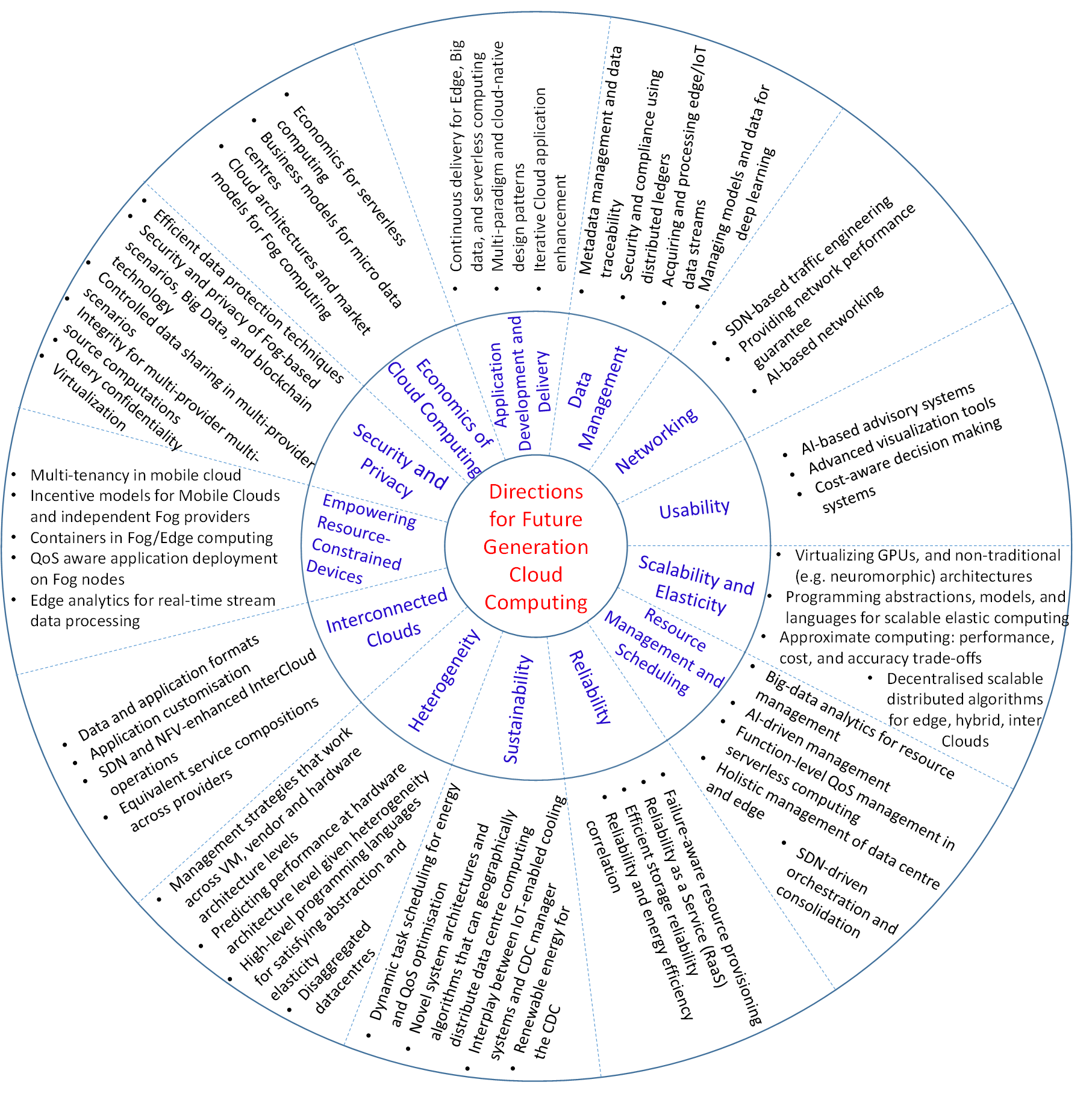}
  \caption{Future research directions in the Cloud computing horizon}
  \label{fig:CloudFutureResearch}
\end{figure}

As can be observed from the emerging trends and proposed future research directions (summarised in the outer ring of Figure~\ref{fig:CloudFutureResearch}), there will be significant developments across all the service models (IaaS, PaaS and IaaS) of Cloud computing.

In the IaaS there is scope for heterogeneous hardware such as CPUs and accelerators (e.g. GPUs and TPUs) and special purpose Clouds for specific applications (e.g. HPC and deep learning). The future generation Clouds should also be ready to embrace the non-traditional architectures, such as neuromorphic, quantum computing, adiabatic, nanocomputing etc.  Moreover, emerging trends such as containerisation, SDN and Fog/Edge computing are going to expand the research scope of IaaS by leaps and bounds. Solutions for addressing sustainability of CDC through utilisation of renewable energy and IoT-enabled cooling systems are also discussed. There is also scope for emerging trends in IaaS, such as disaggregated data centres where resources required for the computational tasks such as CPU, memory and storage, will be built as stand-alone resource blades, which will allow faster and ideal resource provisioning to satisfy different QoS requirements of Cloud based applications. The future research directions proposed for addressing the scalability, resource management and scheduling, heterogeneity, interconnected Clouds and networking challenges, should enable realising such comprehensive IaaS offered by the Clouds.
 
Similarly, PaaS should see significant advancements through future research directions in resource management and scheduling. The need for programming abstractions, models, languages and systems supporting scalable elastic computing and seamless use of heterogeneous resources are proposed leading to energy-efficiency, minimised application engineering cost, better portability and guaranteed level of reliability and performance. It is also foreseeable that the ongoing interest for ML, deep learning, and AI applications will help in dealing with the complexity, heterogeneity, scale and load balancing applications developed through PaaS. Serverless computing is an emerging trend in PaaS, which is a promising area to be explored with significant practical and economic impact. Interesting future directions are proposed such as function-level QoS management and economics for serverless computing. In addition, future research directions for data management and analytics are also discussed in detail along with security, leading to interesting applications with platform support such as edge analytics for real-time stream data processing, from the IoT and smart cities domains.
 
SaaS should mainly see advances from the application development and delivery, and usability of Cloud services. Translucent programming models, languages, and APIs will be needed to enable tackling the complexity of application development while permitting control of application delivery to future-generation Clouds. A variety of agile delivery tools and Cloud standards (e.g., TOSCA) are increasingly being adopted during Cloud application development. The future research should focus at how to continuously monitor and iteratively evolve the design and quality of Cloud applications. It is also suggested to extend DevOps methods and define novel programming abstractions to include within existing software development and delivery methodologies, a support for IoT, Edge computing, Big Data, and serverless computing. Focus should also be at developing effective Cloud design patterns and development of formalisms to describe the workloads and workflows that the application processes, and their requirements in terms of performance, reliability, and security are strongly encouraged. It is also interesting to see that even though the technologies have matured, certain domains such as mobile Cloud, still have adaptability issues. Novel incentive mechanisms are required for mobile Cloud adaptability as well as for designing Fog architectures.

Future research should thus explore Cloud architectures and market models that embrace uncertainties and provide continuous ``win-win'' resolutions, for all the participants including providers, users and intermediaries, both from the Return On Investment (ROI) and satisfying SLA perspectives.

\section{Summary and Conclusions}
\label{sec:summary}

The Cloud computing paradigm has revolutionised the computer science horizon during the past decade and enabled emergence of computing as the fifth utility. It has emerged as the backbone of modern economy by offering subscription-based services anytime, anywhere following a pay-as-you-go model. Thus, Cloud computing has enabled new businesses to be established in a shorter amount of time, has facilitated the expansion of enterprises across the globe, has accelerated the pace of scientific progress, and has led to the creation of various models of computation for pervasive and ubiquitous applications, among other benefits.
 
However, the next decade will bring about significant new requirements, from large-scale heterogeneous IoT and sensor networks producing very large data streams to store, manage, and analyse, to energy- and cost-aware personalised computing services that must adapt to a plethora of hardware devices while optimising for multiple criteria including application-level QoS constraints and economic restrictions. These requirements will be posing several new challenges in Cloud computing and will be creating the need for new approaches and research strategies, and force us to re-evaluate the models that were already developed to address the issues such as scalability, resource provisioning, and security. 
 
This comprehensive manifesto brought the advancements together and proposed the challenges still to be addressed in realising the future generation Cloud computing. In the process, the manifesto identified the current major challenges in Cloud computing domain and summarised the state-of-the-art along with the limitations. The manifesto also discussed the emerging trends and impact areas that further drive these Cloud computing challenges. Having identified these open issues, the manifesto then offered comprehensive future research directions in the Cloud computing horizon for the next decade. The discussed research directions show a promising and exciting future for the Cloud computing field both technically and economically, and the manifesto calls the community for action in addressing them.

\section*{Acknowledgement}
We thank anonymous reviewers, Sartaj Sahni (Editor-in-Chief) and Antonio Corradi (Associate Editor) for their constructive suggestions and guidance on improving the content and quality of this paper. We also thank Adam Wierman (California Institute of Technology), Shigeru Imai (Rensselaer Polytechnic Institute) and Arash Shaghaghi (University of New South Wales, Sydney) for their comments and suggestions for improving the paper. Regarding funding, G. Casale has been supported by the Horizon 2020 project DICE (644869).

\bibliographystyle{abbrv} 
\bibliography{CloudManifesto}

\end{document}